\documentclass{aa}
\usepackage[varg]{txfonts}
\usepackage{natbib}
\bibpunct{(}{)}{;}{a}{}{,}
\usepackage{footnote}
\usepackage{longtable}
\usepackage{lscape}
\usepackage{verbatim}
\usepackage{makecell}
\usepackage{multirow}
\usepackage{booktabs}
\usepackage{soul}
\usepackage{gensymb}
\usepackage{graphicx}
\usepackage{placeins}
\usepackage{ragged2e}
\usepackage{hyperref}
\usepackage{array}
\usepackage{subcaption}
\usepackage{textcomp}
\usepackage{float}
\usepackage{lastpage}
\usepackage{caption}

\newcommand\rurl[1]{%
  \href{http://#1}{\nolinkurl{#1}}%
}

\makeatletter
\renewcommand*\aa@pageof{, page \thepage{} of \pageref*{LastPage}}
\makeatother

\begin{document}

\title{Variability in hot sub-luminous stars and binaries: Machine-learning analysis of \textit{Gaia} DR3 multi-epoch photometry}

   \author{P. Ranaivomanana 
          \inst{1,2} \and M. Uzundag \inst{2} \and C. Johnston \inst{1,2,3} \and P.J. Groot \inst{1,4,5,6} \and T. Kupfer \inst{7,8} \and C. Aerts \inst{1,2,9}
          } 
   \institute{
   Department of Astrophysics/IMAPP, Radboud University, P.O.Box 9010, 6500 GL Nijmegen, The Netherlands\\
   \email{princy.ranaivomanana@ru.nl}
   \and 
   Instituut voor Sterrenkunde, KU Leuven, Celestijnenlaan 200D, 3001 Leuven, Belgium \and
   Max-Planck-Institut für Astrophysik, Karl-Schwarzschild-Straße 1, 85741 Garching bei München,Germany \and
   Department of Astronomy, University of Cape Town, Private Bag X3, Rondebosch, 7701, South Africa \and South African Astronomical Observatory, P.O. Box 9, Observatory, 7935, South Africa \and  The Inter-University Institute for Data Intensive Astronomy, University of Cape Town, Private Bag X3, Rondebosch, 7701, South Africa \and
   Hamburger Sternwarte, University of Hamburg, Gojenbergsweg 112, 21029 Hamburg, Germany \and 
   Texas Tech University, Department of Physics \& Astronomy, Box 41051, 79409, Lubbock, TX, USA \and
   Max Planck Institute for Astronomy, Königstuhl 17, 69117 Heidelberg, Germany 
    }
 
   \date{Received 30 September, 2024 / accepted 3 December 2024}

 
  \abstract
{Hot sub-luminous stars represent a population of stripped and evolved red giants that is located on the extreme horizontal branch. Since they exhibit a wide range of variability due to pulsations or binary interactions, it is crucial to unveil their intrinsic and extrinsic variability to understand the physical processes of their formation. In the Hertzsprung-Russell diagram, they overlap with interacting binaries such as cataclysmic variables (CVs).}
   {
By leveraging the most recent clustering algorithm tools, we investigate the variability of 1,576 candidate hot subdwarf variables using comprehensive data from {\it Gaia} DR3 multi-epoch photometry and Transiting Exoplanet Survey Satellite (TESS) observations. 
   }
{
We present a novel approach that uses the  t-distributed stochastic neighbour embedding and the uniform manifold approximation and
projection dimensionality reduction algorithms to facilitate the identification and classification of different populations of variable hot subdwarfs and CVs in a large dataset. In addition to the publicly available {\it Gaia} time-series statistics table, we adopted additional statistical features that enhanced the performance of the algorithms.
}
{ The clustering results led to the identification of 85 new hot subdwarf variables based on {\it Gaia} and TESS light curves and of 108 new variables based on {\it Gaia} light curves alone, including reflection-effect systems, HW Vir, ellipsoidal variables, and high-amplitude pulsating variables. A significant number of known CVs (140) distinctively cluster in the 2D feature space among an additional 152 objects that we consider candidates for new CVs. 
   }
{This study paves the way for more efficient and comprehensive analyses of stellar variability from ground- and space-based observations, and for the application of machine-learning classifications of candidate variable stars in large surveys.}


\titlerunning{Variability of hot sub-luminous stars and binaries}
\authorrunning{Ranaivomanana et al.}
\maketitle
%
\section{Introduction}
Hot sub-luminous stars are hot and compact evolved low-mass stars that are located on the extreme horizontal branch, between the main sequence (MS) and the white dwarf sequence  \citep{Heber2009,Heber2016}. In a Hertzsprung-Russell diagram (HRD), they occupy B and O spectral types and form the population of hot subdwarf B (sdB) and O (sdO) stars. A recent study of a 500 pc volume-limited sample of hot sub-luminous stars reported that they are dominated by the sdB population ($\sim 60\%$; \citealt{Dawson2024}). Most of this population are thought to have a canonical core mass of 0.47 $M_{\odot}$ and thin hydrogen layers ($\sim 10^{-4} -  10^{-2}\, \rm M_{\odot}$; \citealt{Saffer1994,Brassard2001}). Their thin envelope mass suggests that sdBs are the remnant cores of low-mass red giant stars that were stripped through binary interactions, which introduced a different evolutionary path than for normal horizontal branch stars. This envelope mass prevents them from supporting H-shell burning. After depletion of helium in the sdB cores, on a timescale of $\sim 10^8\, \rm yr$ \citep{Dorman1993,Ostrowski2021}, they first become sdOs and then evolve to the white dwarf cooling stage. 

Evolutionary calculations showed that sdB progenitors likely underwent binary interactions \citep{Han2002,Han2003}, including common-envelope ejection (CEE; for short-period binaries with a period of 0.1$-$10 days), stable Roche-lobe overflow (RLOF; for long-period or composite binaries with periods of 450$-$1600 days; \citealt{Vos2020}), and mergers (e.g. He white dwarf + He white dwarf; \citealt{Webbink1984}). Observational studies corroborated this \citep{Pesoli2020}, and multiple studies reported a significant fraction of hot subdwarfs in binary systems, either in close binaries with a MS or white dwarf companion (e.g. \citealt{Geier2022,Schaffenroth2022, Schaffenroth2023}), or in wide binaries with cool MS companions (e.g. \citealt{Deca2012,vos2019,Vos2020}). This diversity makes them an excellent population for studying binary star evolution. In addition, a broad range of unseen companions have been confirmed to exist in hot subdwarfs, such as low-mass MS stars (dM), brown dwarfs, and white dwarf companions \citep{Kupfer2015,Geier2010, Geier2022, Geier2023} through the project called Massive Unseen Companions to Hot Faint Underluminous Stars from the Sloan Digital Sky Survey (MUCHFUSS). The existence of these companions and their nature is often shown by the behaviour of photometric light curves of the hot subdwarfs, such as ellipsoidal variability for white dwarf companions and reflection effects for low-mass companions \citep{Schaffenroth2022, Barlow2022}. 

A population of hot subdwarfs was also found to exhibit pulsations, and asteroseismology was used to study their structure and evolution (e.g. \citealt{Charpinet2010,VanGrootel2010,Reed2020,Sahoo2020,Silvotti2022, Krzesinski2022,Uzundag2021, Uzundag2023,Uzundag2024}). While the mechanism for exciting pulsations in subdwarfs is thought to be understood \citep[i.e. the $\kappa$-mechanism operating on the Fe opacity bump; ][]{Charpinet1997,Fontaine2003}, it is unclear why only a handful of subdwarfs are observed to pulsate while most do not. Theoretical work has demonstrated that atomic diffusion is required, but it is unclear whether other aspects such as the evolution history of the binary also play a role \citep{Hu2008,Hu2011,Bloemen2014}.

It is essential to increase the detection of new variable hot subdwarfs to enable a robust characterisation of their variability and to improve our understanding of these stars. In addition to spectroscopic identifications of hot subdwarfs (e.g. \citealt{Luo2019,Lei2020,Lei2023}), which are often observationally expensive, previous efforts to identify candidate hot subdwarfs were made mainly based on their locations in the colour-magnitude diagram and proper motion selection criteria \citep{Geier2019,Geier2020} using {\it Gaia} DR2 observations \citep{GaiaCollab2018}. Following similar steps, \cite{Culpan2022} compiled a large catalogue of more than 60,000 confirmed and candidate hot subdwarfs observed from {\it Gaia} EDR3 data \citep{Gaia_collab2021}. These selections are frequently affected by contamination from low-mass MS stars, cataclysmic variables (CVs), and white dwarfs \citep{Geier2019,Culpan2022, Barlow2022}. Given this contamination, it is critical for target selections to develop an effective framework to separate hot subdwarfs from other populations of blue objects in the HR diagram and characterise their variability in multiple time-domain surveys.

The interest in developing machine-learning algorithms to automate the variability search and characterisation of time-series data in time-domain astronomy has been strong (e.g.\citealt{Kim2021,Cui2022,Eyer2023, Monsalves2024}) due to the growing volume of data generated by large surveys, such as the All Sky Automated Survey (ASAS; \citealt{Pojmanski2002}), the \textit{Zwicky} Transient Facility (ZTF; \citealt{Bellm2019}), and the {\it Gaia} mission \citep{GaiaCollab2023}. As the majority of these algorithms either depend on a particular survey (e.g. based in space or on the ground) or are task-oriented (e.g. a planet transit detection), their application is often limited to a certain number of specific cases and goals.

To remedy this, we present a machine-learning framework for identifying variable hot subdwarfs and CVs based on photometric time series alone. Our methods can be broadly applied to any photometric data, such as those from the BlackGEM \citep{Groot2024}, the Gravitational-wave Optical Transient Observer (GOTO; \citealt{Steeghs2022}), and the Legacy Survey of Space and Time (VRO/LSST; \citealt{Ivezic2019}) missions. The structure of this paper is as follows: In Sect. \ref{sec:data_and_method} we describe the data and methods. This is followed by feature engineering and the cluster analysis in Sect. \ref{sec:feature_engineering}. The results of the variability classification are provided and discussed in Sect. \ref{sec:results}. Our conclusion and future prospects are presented in Sect. \ref{sec:conclusion}.

\section{Data and methods}\label{sec:data_and_method}
\subsection{{\it Gaia} observations}\label{sec:Gaia_obs}
The precise astrometric and photometric measurements provided by {\it Gaia} significantly boost the identification of the population of candidate hot subdwarfs in the colour-magnitude diagram. \cite{Culpan2022} compiled a catalogue of 61,585 candidate hot subdwarfs based on colour, absolute magnitude, and reduced proper motion selection criteria in {\it Gaia} EDR3 \citep{Gaia_collab2021}, which served as the basis of this work. The release of {\it Gaia} DR3 multi-epoch photometry \citep{Eyer2023} allowed us to cross-match this catalogue to find candidates with available light curves and further study their variability.
This resulted in 2114 objects with available epoch photometry using the {\it Gaia} flag has\_epoch\_photometry=True. The remaining 59,471 objects were excluded from the analysis because no {\it Gaia} light curves are available for them.

Using the {\it Gaia} datalink service and the $astroquery.Gaia$ package \citep{Ginsburg2019}, we extracted the light curves of these objects in the three {\it Gaia} filter bands (G, BP, and RP). Before we searched for periodicity, we preliminarily assessed the quality. First, we retained objects with reliable parallax measurements (parallax\_over\_error > 5). Second, the {\it Gaia} boolean quality flag \texttt{reject\_by\_variability} was used to remove data points rejected by the {\it Gaia} variability pipeline \citep{Eyer2023}, and then objects with at least 25 observations in any of the three band light curves were selected, following the minimum number of observations suggested by \cite{Morales-Rueda2006} for detecting stellar variability. For the {\it Gaia} astrometric quality control, known as the re-normalised unit weight error (RUWE), RUWE$<$7 was adapted as a substantial number of spectroscopically identified hot subdwarfs were observed to exceed the recommended  RUWE$<$1.4 limit up to RUWE=7 (see \cite{Dawson2024} for more details). These selections resulted in 1,682 light curves that were ready for analysis. Their {\it Gaia} G-band light curves have a typical median signal-to-noise (S/N) ratio estimate (standard deviation of the magnitudes over the rms of the magnitude uncertainties) of 3.5 and a median number of observations of about 40, as well as a median magnitude of $\sim$15 mag.

\subsection{Frequency analysis}\label{sec:freq_analysis}
The population of hot subdwarfs hosts diverse types of variability, including pulsating variables and eclipsing binaries, from close- to wide-binary systems. Therefore, their variability exhibits a wide range of timescales from minutes to months and of morphologies from sharp eclipses to sinusoidal pulsations. Following the success of our frequency-search algorithms in finding dominant frequencies in multi-band, heteroscedastic, and irregularly sampled light curves of candidate hot subdwarfs from the MeerLICHT telescope \citep{Ranaivomanana2023}, we applied the same approach to search for periodicity in {\it Gaia} light curves. In brief, this method combines Fourier-based calculations, namely the generalised Lomb-Scargle periodogram, and phase-dispersion measurements, known as Lafler-Kinman statistics, to alleviate the effects of noise and data gaps in a periodogram. This hybrid approach is referred to as the $\Psi-$static \citep{Saha2017}, where the notation $\Psi$ is used to represent the periodogram throughout this work. 

In the frequency grid search, the search was performed from zero up to 360 day$^{-1}$ according to the Nyquist-frequency of the 2-minute cadence of the Transiting Exoplanet Survey Satellite (TESS; \citealt{Ricker2015}) short-cadence observations, which were used for comparison with the {\it Gaia} variability in Sect. \ref{sec:results}. The frequency step was finely tuned and was defined as the inverse of the total time base divided by an oversampling factor of 10, following results in the literature that showed that this value is appropriate to ensure that no dominant frequency peaks are missed and to prevent a poor period estimation, which would occur if its value were taken too low \citep{Vanderplas2018,Schwarzenberg-Czerny1996}. In addition, the dominant frequency we found was further optimised by fine-tuning the frequency step with an oversampling factor of 100. This was only done in a small frequency window around the dominant frequency, where a frequency window size ten times larger than the original frequency step was used on either side of the peak. 

\subsection{Uncertainties in the frequency estimates}

The uncertainties in the dominant frequencies were estimated by adopting a Monte Carlo approach, where the frequency algorithm was run 1000 times. The standard deviation of the dominant frequencies was taken as an estimate of the frequency uncertainty. Each iteration consisted of 1) drawing a sample from a normal distribution with zero mean and a width of the magnitude errors per observation, and 2) creating a new light curve by adding the sample to the original light curve. The magnitude errors in the original light curves were kept in the new light curves. The iterations finally consisted of 3) running the algorithm on the new light curve using the same fine-tuned frequency window as in the frequency optimisation. Due to the finite sampling step in the frequency grid, each iteration could result in the same identified dominant frequency. To mitigate this, the frequency grids were shifted by 1/1000th of the frequency step for each of the 1000 iterations to ensure that the frequency search was not confined to the same frequency peak in each iteration. 

\section{Feature engineering}\label{sec:feature_engineering}
\subsection{Variability analysis}\label{sec:var_analysis}
After we computed the dominant frequencies for all candidates, a robust, unbiased method was required to determine the significance of the peaks and measure the reliability of the variability. Although the false-alarm probability (FAP; \citealt{Scargle1982,Baluev2008}) was frequently used in the literature to measure the significance of the frequency peak, it is poorly adapted to variables in the high-frequency domain and in the case of signals with red noise \citep{Vanderplas2018}. Additionally, the interpretation of the FAP becomes complex in our case, where two independent periodograms were combined in the hybrid approach. Therefore, we addressed this by exploring machine-learning clustering algorithms to distinguish candidates with different significance levels and variability. 

We explored various summary statistics that are capable of unveiling the fidelity of the frequency peaks and the variability in the {\it Gaia} time series. It was necessary to extract these parameters from the data to work with the clustering algorithms described in the next sections. First, we extracted the {\it Gaia} variability summary statistics table\footnote{\url{https://doi.org/10.17876/Gaia/dr.3/92}}, which consists of statistical parameters (54 in total, excluding boolean parameters and object IDs) that were computed using the {\it Gaia} DR3 time series \citep{Eyer2023}. Second, after normalising the maximum amplitude in the $\Psi-$periodogram to one, we computed additional statistical features (24 features) that were specifically designed to help us define the significance of the peak, such as the 95th percentile of the amplitude for the 100 peaks with the highest amplitude, the 99th percentile of the amplitudes for the full spectrum, and the number of frequencies with amplitudes above 0.5. These features were found to be useful for distinguishing objects with a clear variability, as we discuss in Sect. \ref{sec:feature_selection}. We obtained 84 features in total (see Table \ref{tab:all_features}) when we combined these features with the {\it Gaia} statistics table and another six parameters from the {\it Gaia} DR3 source database \citep{GaiaCollab2023}, such as BP$-$RP, parallax, and RUWE. Entries with missing values were removed from the table, which left 1,576 final candidates out of the 1,682 objects.
\begin{figure*}
  \centering
\includegraphics[width=0.95\linewidth]{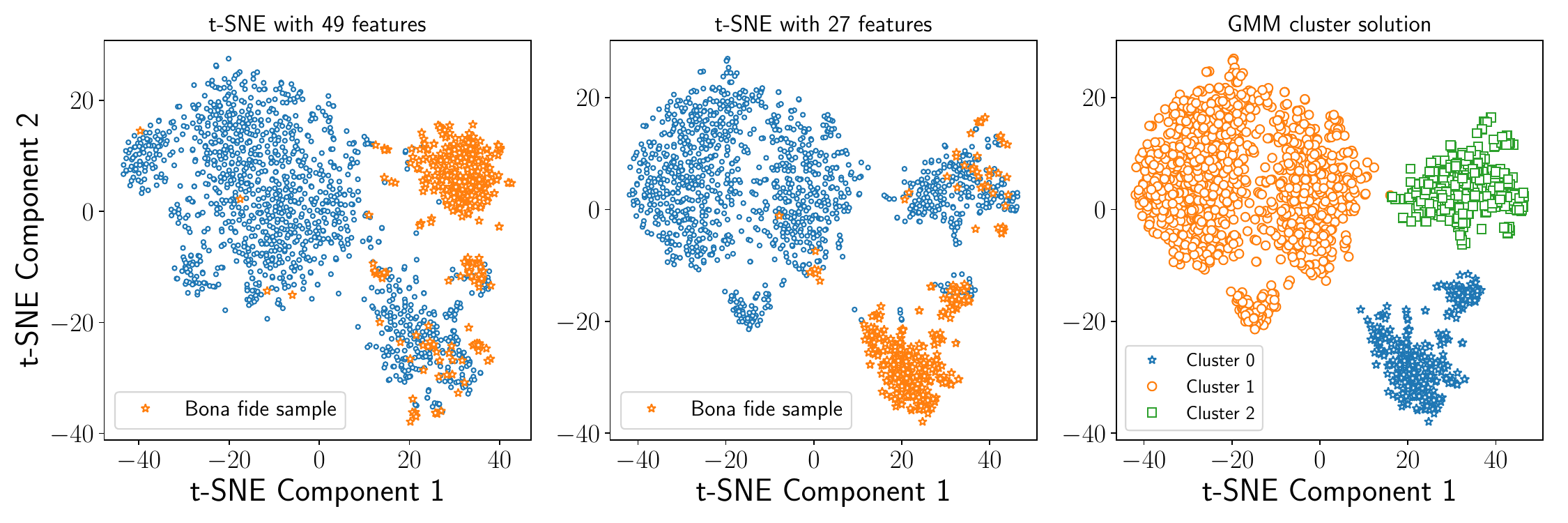}\\
\includegraphics[width=0.95\linewidth]{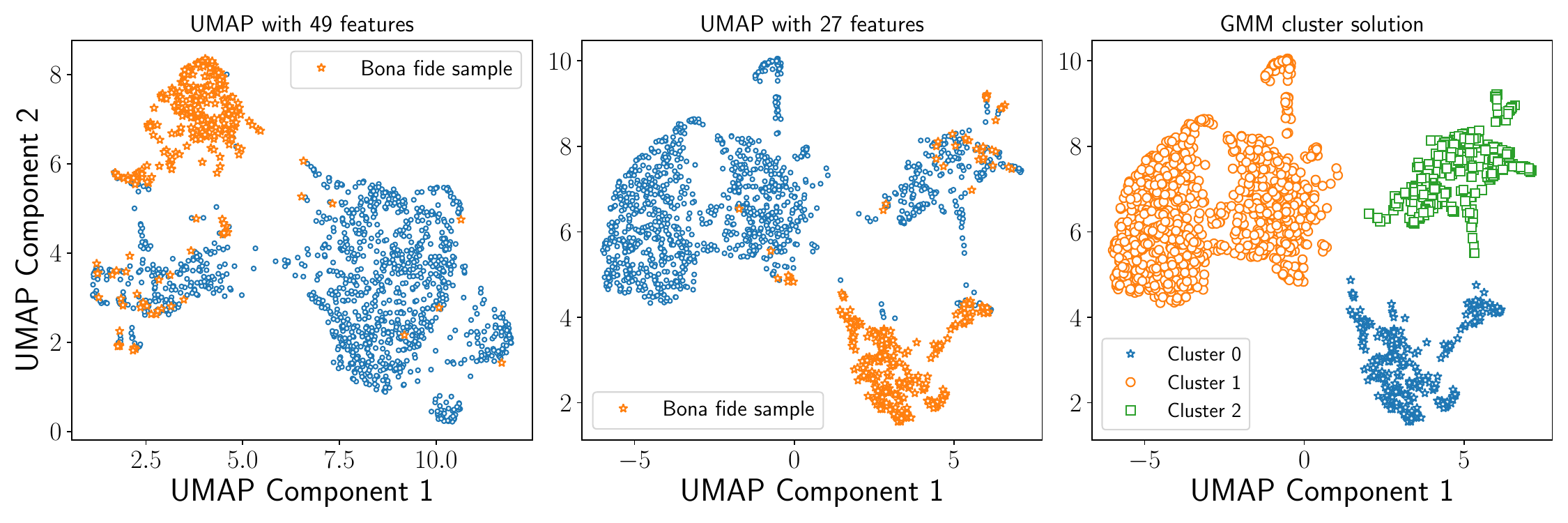}
\caption{\small Clustering results using the t-SNE (top panels) and UMAP (bottom panels) dimensionality reduction algorithms and clustering labels (right panels) from the Gaussian mixture model (GMM). The left and middle panels represent the 2D components using 49 and 27 features, respectively. The open orange stars in these panels correspond to the manually selected objects with clear variabilities.
}\label{fig:clustering}
\end{figure*}
\subsection{Dimensionality reduction}\label{sec:feature_selection}
The next step was to transform these features into a lower-dimensional space such that we were able to visualise and identify possible clusters. This was done by applying dimensionality reduction techniques to our data, which convert high-dimensional features into a 2D feature space. It is common practice to reduce the dimensionality in machine learning, and it has been used extensively in astronomy to visualise and interpret data \citep{Kao2024,Liao2024,Pantoja2022}. We explored two non-linear dimensionality reduction techniques: the t-distributed stochastic neighbour embedding (t-SNE; \citep{Vandermaaten2008}) and the uniform manifold approximation and projection (UMAP; \citep{McInnes2018}). These were chosen over other techniques such as the principal component analysis (PCA) because they are able to find non-linear structures in data and are straightforward to implement. 

Since our data had more than 80 features, it was important to remove highly correlated features that might lead to noise in the visualisation \citep{kuhn2019}. This also helped the algorithms, notably t-SNE, to efficiently map the high-dimensional to low-dimensional space. To identify correlated features, we calculated the Pearson correlation coefficient between all pair-wise combinations of features and excluded one feature from each pair for correlation values above 0.95. This reduced the number of features to 49, which is also recommended\footnote{\url{https://scikit-learn.org/stable/modules/generated/sklearn.manifold.TSNE.html}} ($\sim$ 50) to efficiently optimise the t-SNE algorithm.

We further ranked the features using a random forest algorithm \citep{breiman2001}, which is a commonly used technique for obtaining relative feature importance scores (e.g. \citealt{Richards2012}). The importance score of each feature is determined based on its ability to split the data into pure nodes (nodes with instances belonging to the same class) in the individual decision trees of the random forest model (see \cite{breiman2001} for more details). At this stage, the sole purpose was to obtain the feature scores. Therefore, the default random forest model hyper-parameters (e.g. the number of estimators) were used to fit the data. Upon model fitting, we obtained the relative importance scores of each feature. These scores were used to optimise the t-SNE and UMAP algorithms in Sects. \ref{sec:tsne} and \ref{sec:umap}.

We also manually labelled each object based on their phase-folded diagrams, where objects that exhibited an obvious variability were labelled as 0, and those with an ambiguous variability were labelled as 1. These labels were used when fitting the random forest algorithm. In addition, labelling the data allowed us to examine the clustering performances and to visualise the physical or statistical distribution of each class (e.g. period distribution) in the clusters shown in Fig. \ref{fig:clustering}, which we discuss further throughout the paper. 
\begin{figure*}
  \centering
\begin{tabular}{cc}
 \includegraphics[width=0.45\linewidth]{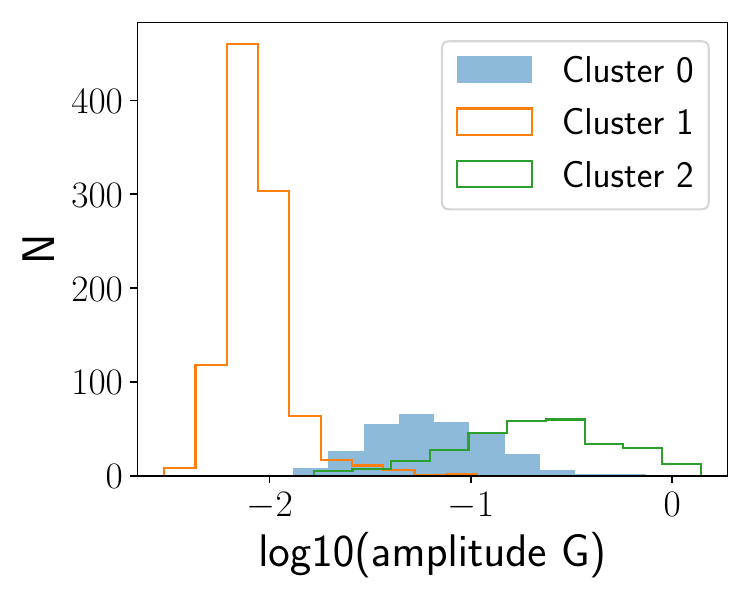}  
     &  \includegraphics[width=0.5\linewidth]{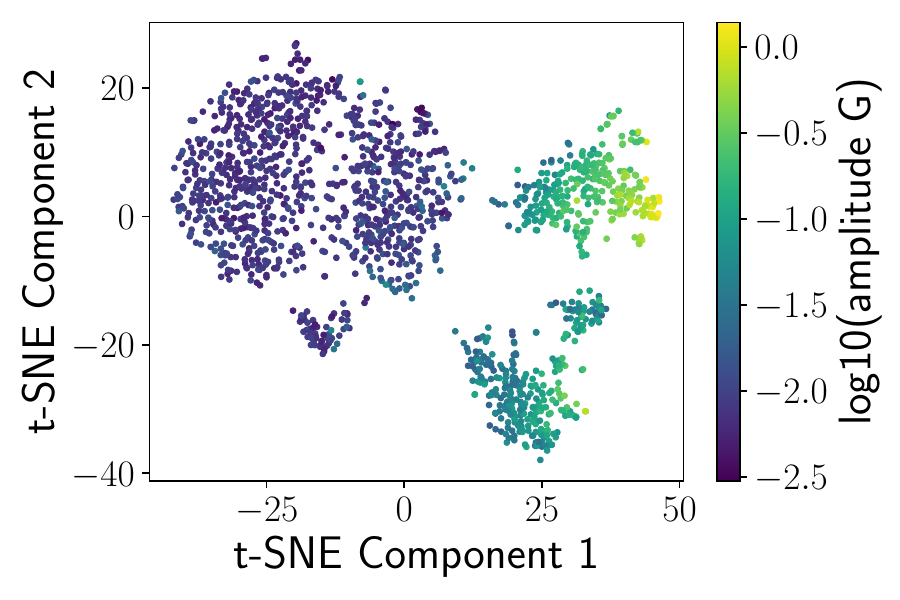}
\end{tabular}
\caption{\small Amplitude distribution of each cluster in the {\it Gaia} G band (left panel) and gradient of the variability amplitude in the G band across the t-SNE components (right panel).}\label{fig:cls_amp_gradient} 
\end{figure*}
\subsubsection{Dimensionality reduction with t-SNE}\label{sec:tsne}
We implemented the \texttt{TSNE} module from the \texttt{scikit-learn} Python library \citep{scikit-learn}, where two crucial parameters, namely perplexity and learning rate, were optimised, while the other parameters were kept to their default values. The perplexity can be seen as a tuning parameter that measures the effective number of nearest neighbours to be considered to construct the low-dimensional embedding. Before running the t-SNE algorithm, we first scaled each feature to have a zero mean and unit standard deviation, which helped the algorithm to be more efficient in finding structures in the data. The optimised values of the two parameters are perplexity = 50 and learning rate = 600. With these settings and the 49 features, Fig. \ref{fig:clustering} shows the transformed low-dimensional projections, where we can identify three main clusters, namely cluster 0, cluster 1, and cluster 2. These are discussed in more detail in Sect. \ref{sec:cluster_analysis}. The open orange stars in the left and middle panels of Fig. \ref{fig:clustering} represent the objects we labelled manually, most of which belong to one cluster.
To label these clusters, we fit the 2D projection data to a Gaussian mixture model (implemented in scikit-learn) with three mixture components. The advantage of using this model is that it provides the probability of each object to belong to a cluster. The quality of the class labels predicted by the Gaussian mixture model was evaluated using the so-called silhouette score \citep{Rousseeuw1987}, in addition to a visual inspection of the graphical output. This evaluation metric compares how well data points match their designated cluster to other clusters. We obtained a silhouette score of 0.535, which is generally considered to indicate a reasonable clustering solution (i.e. $>0.5$; \citealt{Rousseeuw1987}). We further improved this by iteratively removing the least important features from the importance scores computed above that might cause noise in the low-dimensional representation. In other words, we stopped the iterative process when no further improvements were visually detected in the output clusters and in the silhouette score. This resulted in 27 features with a silhouette score of 0.567. The t-SNE 2-D representation of this result is shown in Fig. \ref{fig:clustering} together with the Gaussian mixture clustering solution. These 27 features are described in Table \ref{tab:feature_ranking_manual_lab} and are used throughout the rest of the analysis. 

The dominant features for the manually labelled objects include the 95th percentile of the first 100 frequency peaks, the number of peaks above 0.5 of the normalised $\Psi$ periodogram, and the 99th percentile of all periodogram peaks. However, they do not imply that these top features alone can explain the separation of the three clusters in the 2D feature space; it only means that their importance scores are higher than those for the rest of the features, as shown in Fig. \ref{fig:features_importance_score}. As previously mentioned, the aim of dimensionality reduction algorithms is to build new low-dimensional features from linear or non-linear combinations of high-dimensional features while preserving as much of the original information as possible. Since the low-dimensional features are mixtures of the original ones, we cannot conclude from the 2D representation that a specific or a group of a few features cause the distinction of the clusters.
 
\subsubsection{Dimensionality reduction with UMAP}\label{sec:umap}
Similar to t-SNE, UMAP \citep{McInnes2018} is a non-linear algorithm for high-dimensional data visualisation, except that its approach to dimensionality reduction is grounded in manifold theory and topological data analysis rather than probabilistic modelling as in t-SNE. The UMAP algorithm is implemented in the umap-learn Python package \citep{umap-software2018}. We ran the UMAP algorithm with its default parameter values and the selected 27 features in Sect. \ref{sec:tsne}, which already resulted in reasonable silhouette score values and a distinctive visualisation (Fig. \ref{fig:clustering}). The same features as obtained from t-SNE were also used when running the UMAP algorithm to show that both algorithms output the same results using the same features, and to obtain meaningful clustering results. The obtained silhouette scores are very similar for the 27 features (0.597) and 49 features (0.599). The cluster labels were again obtained from the Gaussian mixture model. We identified three main clusters similar to those found with the t-SNE algorithm, which confirms the existence of these clusters in our data. The next section compares the results from the two algorithms.
\begin{figure*}
  \centering
\begin{tabular}{ccc}

\includegraphics[width=0.32\linewidth]{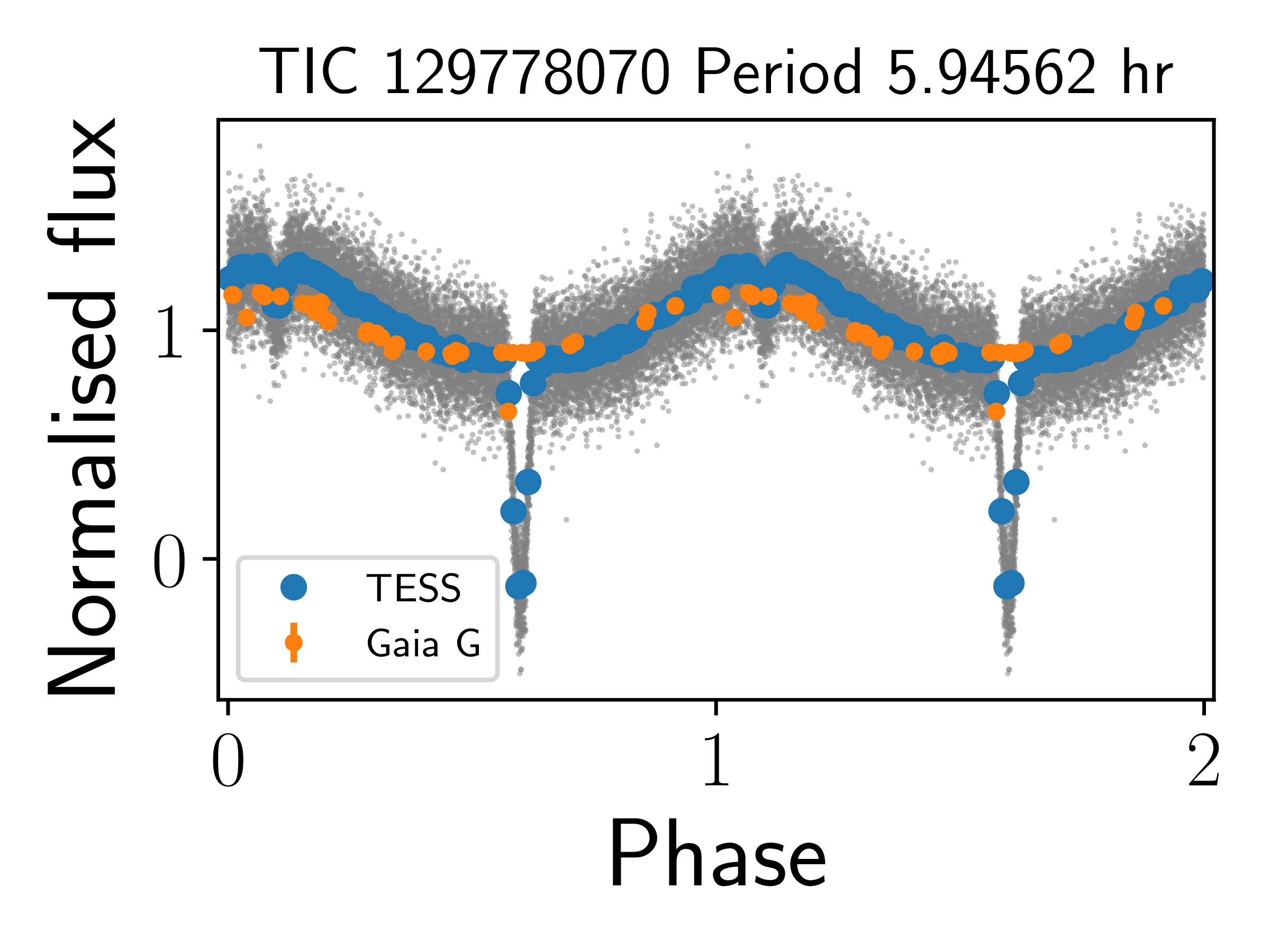}&
\includegraphics[width=0.32\linewidth]{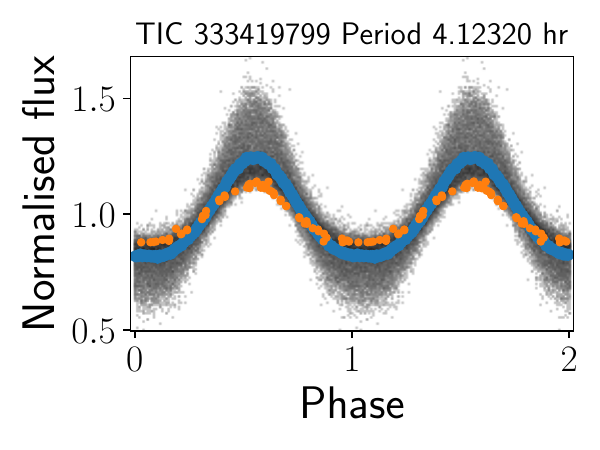}&
\includegraphics[width=0.32\linewidth]{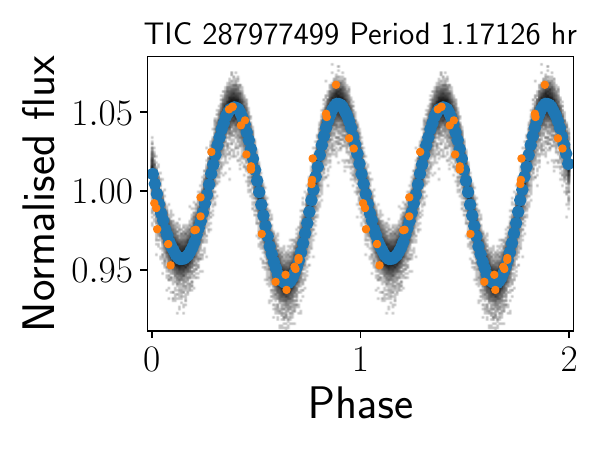}
\end{tabular}
\caption{\small Three examples of new variable hot subdwarfs identified in this work. The light curves are phase-folded to the same periods and reference epochs. The three objects correspond to an HW Vir (left), a reflection effect (middle), and an ellipsoidal system (right). The light curves on the right are phased to twice the period to highlight the ellipsoidal variation. The blue lines represent the binned phase of the TESS light curves (grey data points), and the orange data points correspond to {\it Gaia} light curves.
}\label{fig:sample_var}
\end{figure*}
\subsection{Cluster analysis and candidate selection}\label{sec:cluster_analysis}
It is worth examining whether the three clusters found by t-SNE and UMAP represent the same objects. Of the t-SNE and UMAP components, 290 and 297 objects are part of cluster 0; 990 and 991 objects in cluster 1; and 296 and 288 objects in cluster 2, respectively. Therefore, we cross-matched the objects in the three clusters from both algorithms and found a total number of 1563/1576 matches ($\sim 99\%$): 289, 988, and 286 matches from cluster 0, cluster 1, and cluster 2, respectively. This shows that the two clustering approaches are highly consistent. We examined the 13 mismatched objects because the clustering results for t-SNE and UMAP matched for 1,563 out of 1,576 objects. Eight of these 13 objects belong to cluster 0 in UMAP and to cluster 2 in t-SNE. These objects exhibit large peak-to-peak magnitudes in the {\it Gaia} G band, with variations of at least 0.5 mag. In the 2D t-SNE plot, they are located near the border of cluster 2, close to cluster 0, which may explain the mismatch in the cluster labels between UMAP and t-SNE for these objects. The remaining 5 of the 13 objects either appear in cluster 1 in UMAP and cluster 2 in t-SNE, or vice versa, and they are similarly positioned at the borders of each cluster. We did not observe any peculiar objects in addition to these cases.

As our primary goal was to identify objects with significant and clear variability among the clusters, we visually examined the light curves of the objects in each cluster. We observed that the three clusters reflect the clarity of the light-curve variability, which can be translated into the light-curve S/N ratio. More precisely, cluster 1 contains objects with a dubious variability that might be related to light curves with a relatively low S/N ratio; cluster 2 primarily consists of objects with ambiguous light-curve shapes but high variability amplitudes; and cluster 0 is dominated by objects with a clear variability that is associated with high S/N ratio light curves. Some examples of light curves in each cluster are shown in Fig. \ref{fig:cluster_lc_examples}, where the top panels represent clear variables that are typical for cluster 0, the middle panels correspond to unclear variables found in cluster 1, and the bottom panels consists of high-amplitude ambiguous variables in cluster 2. Since the two algorithms represent mostly the same objects per cluster, we focused our analysis on the clusters from the t-SNE components.

Furthermore, we measured the importance score of each of the 27 features using random forest based on the assigned label for each cluster, as we did with the manually labelled data. The results indicated that the amplitude of the variability in the G band (\texttt{amp\_G}) has the highest feature score, followed by the difference between the highest and lowest values of the G-band light curves (\texttt{range\_mag\_g\_fov}) and the interquartile range of the G-band light curves (\texttt{iqr\_mag\_g\_fov}). The rest of the features are listed according to their importance score in Table \ref{tab:feature_ranking_clusters}. As shown in the left panel of Fig. \ref{fig:cls_amp_gradient}, the distribution of the amplitude in a log space reveals three distributions that support these results. In the same figure, a lower bound of the amplitude is shown at $\sim 20$ mmag for cluster 0. Additionally, the right panel of Fig. \ref{fig:cls_amp_gradient} reveals that the amplitude values gradually increase from low to high values of the t-SNE component 1.

Based on these results, we considered all objects in cluster 0 (290 objects) as potential variable hot subdwarfs, and we discuss their variability in Sect. \ref{sec:hsd_cand}, while objects in cluster 2 were found to be mostly comprised of CVs and are discussed in Sect. \ref{sec:cv}. 
\begin{figure*}
  \centering
\begin{tabular}{ccc}
 \includegraphics[width=0.32\linewidth]{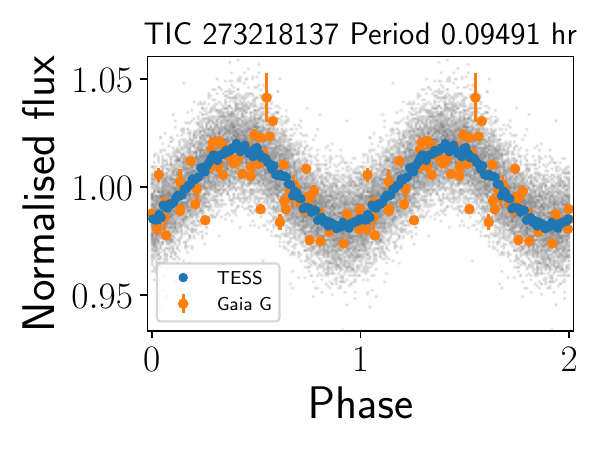}  &  \includegraphics[width=0.32\linewidth]{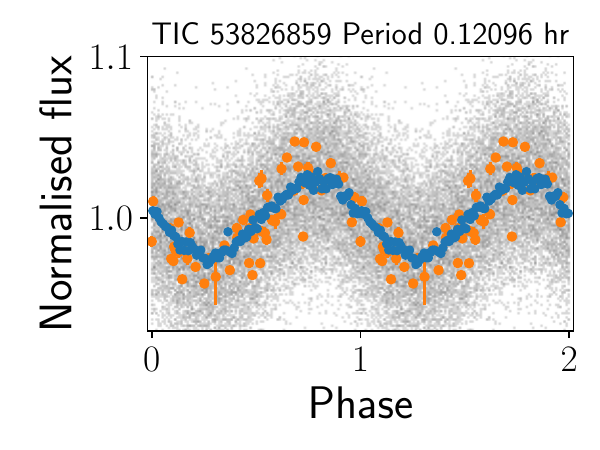}&
 \includegraphics[width=0.32\linewidth]{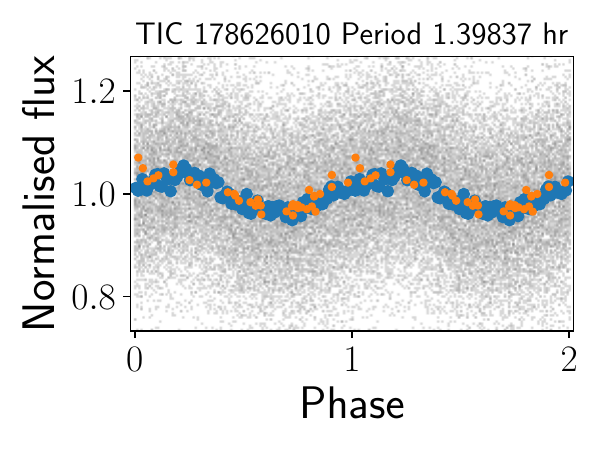}
\end{tabular}
\caption{\small Three pulsating hot subdwarfs observed with {\it Gaia} and TESS. The left and middle panels correspond to known pulsating variables \citep{Krzesinski2022}, and the right panel shows a pulsating variable identified in this work and Krzesinski et al (in prep.). The two light curves are folded to the same periods and reference epochs. The blue lines represent the binned phase of the TESS light curves (grey data points), and the orange data points correspond to {\it Gaia} light curves.
}\label{fig:new_pulsating_var}
\end{figure*}
\begin{figure}
  \centering
\includegraphics[width=0.95\linewidth]{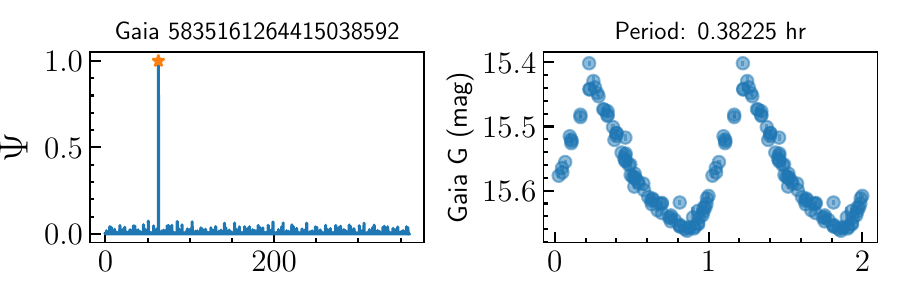}\\
\includegraphics[width=0.95\linewidth]{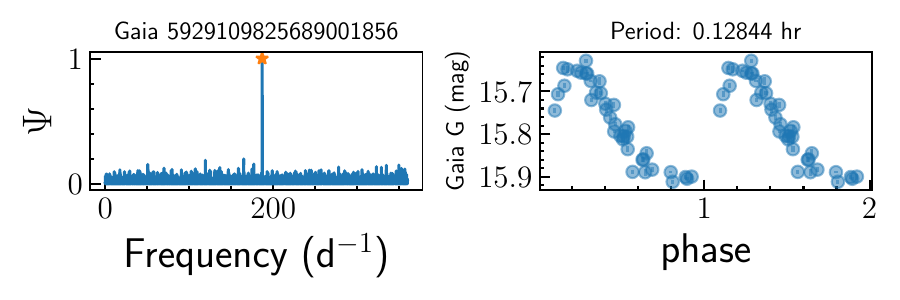}
\caption{\small New high-amplitude pulsating variables observed with {\it Gaia}.
}\label{fig:pulsating_hsd}
\end{figure}
\begin{figure}
    \centering
    \includegraphics[width=0.95\linewidth]{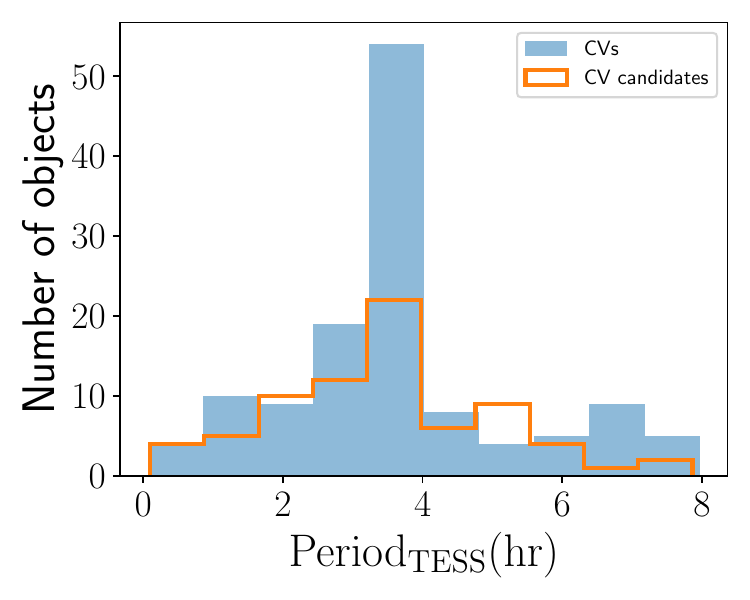}
    \caption{\small Period distribution of the confirmed and candidate CVs in cluster 2.}
    \label{fig:cvs_period_dist}
\end{figure}
\section{Results}\label{sec:results}
\subsection{Hot subdwarf variability classification}\label{sec:hsd_cand}
To confirm the nature of the variations found in the {\it Gaia} light curves, we compared them with those observed by TESS. First, we verified whether any objects in the {\it Gaia} catalogue had light curves in TESS using the \texttt{Lightkurve} Python package \citep{LC_collab2018}. Second, we searched for fast-cadence (20 seconds) and short-cadence (2 minutes) light curves and computed their Lomb-Scargle periodograms. 

The periods found in the {\it Gaia} G band data strongly agree with those obtained by TESS for the objects in cluster 0. The variability types of these objects were thus determined with high confidence. On the other hand, for objects without TESS observations, we are only able to provide a general classification, such as an eclipsing binary or a sinusoidal-like shape class. In order to ensure a homogeneous treatment of the whole sample, we did not rely on TESS data for the results of the frequency analysis. We instead only used the TESS data to improve the fidelity of the classification.
All lists of the candidate classifications are provided in tables A4$-$A10 (see Sect. \ref{sec:data_availability}).

\subsubsection{Variability in the confirmed hot subdwarfs}
We found 78 known variable hot subdwarfs amongst the 290 objects in cluster 0 by cross-matching our data with a catalogue of spectroscopically identified hot subdwarfs and known variable hot subdwarfs from the literature \citep{Schaffenroth2019,Schaffenroth2022,Schaffenroth2023,Culpan2022,Barlow2022,Lei2023, Dawson2024}. Most of them (66/78) were identified from the compiled catalogue of 6,616 known hot subdwarfs \cite{Culpan2022}, and 63/78 have short- or fast-cadence TESS ligh tcurves. Based on the {\it Gaia} and TESS light curves, we found 32 reflection-effect systems, 19 HW Vir systems, 6 pulsating variables, and 6 ellipsoidal variables. The remaining 15/78 systems were classified based solely on the {\it Gaia} three-band light curves, where we found 5 sinusoidal-like light curves that might be associated with reflection-effect systems or ellipsoidal variations or a dominant pulsation mode, 5 eclipsing binaries, and 2 HW Vir systems. Fig. \ref{fig:sample_var} shows examples of new HW Vir (TIC 129778070), reflection effect (TIC 333419799), and ellipsoidal variables (TIC 287977499) systems identified in this work.

\subsubsection{Variability in the candidate hot subdwarfs}
From the unconfirmed hot subdwarfs (212/290), we identified 78 objects with short- and/or fast-cadence TESS light curves. Based on the {\it Gaia} and TESS light curves, we found  42 reflection-effect systems, 21 HW Vir systems, 3 pulsating variables, and 2 ellipsoidal variables. The remaining 134/212 candidate hot subdwarfs were classified based on the {\it Gaia} three-band light curves, where we found 60 sinusoidal-like light curves, 20 HW Vir systems, 14 eclipsing binaries, and 2 potentially pulsating variables. Thirty-eight objects have an unclear variability, which prevented us from labelling them.

\subsubsection{Pulsating hot subdwarfs}
We identified a total of nine already known pulsating variables from the known and candidate hot subdwarfs observed from {\it Gaia} and TESS. Three out of these nine pulsate in the {\it Gaia} and TESS light curves, namely TIC 273218137, TIC 53826859, and TIC 178626010, with a period of 0.09491 hr, 0.12096 hr, and 1.39841 hr, respectively. TIC 273218137 and TIC  53826859 are known pulsating hot subdwarfs from TESS observations \citep{Krzesinski2022}, while TIC 178626010 is a new pulsating variable detected in this work and independently by Krzesinski et al. (in prep). In Fig. \ref{fig:new_pulsating_var}, their {\it Gaia} and TESS light curves are phased to the same periods and reference epochs $t_0$, using the short-cadence light curves for TESS observations. The dominant frequencies found for these two objects are the same in the three {\it Gaia} bands. Therefore, they are candidates for a mode identification from an amplitude ratio analysis \citep{Aerts2023,Fritzewski2024}. Their pulsation frequencies suggest that TIC 273218137 and TIC  53826859 are likely $p-$mode pulsators, and TIC 178626010 pulsates in the $g-$mode regime. The remaining six known pulsating variables have low-amplitude pulsations and higher-amplitude orbital variability in their light curves. In our analysis, we were only able to detect their orbital variability in the {\it Gaia} data. 
\subsubsection{Newly identified pulsating variables}
We identified two unique high-amplitude pulsating objects from {\it Gaia} (Fig. \ref{fig:pulsating_hsd}): {\it Gaia} DR3 5835161264415038592 and {\it Gaia} DR3 5929109825689001856 with G-band peak-to-peak amplitudes of 0.21 mag and 0.25 mag and pulsation periods of 0.38225 hr (22.935 min) and 0.12844 hr (7.706 min), respectively. The BP and RP periods for the two objects are the same as those determined in the G band. Their amplitudes in these bands are as follows: {\it Gaia} DR3 5835161264415038592 has peak-to-peak amplitudes of 0.21 mag and 0.19 mag in the BP and RP bands, respectively. Similarly, {\it Gaia} DR3 5929109825689001856 has peak-to-peak amplitudes of 0.29 mag and 0.23 mag in the BP and RP bands, respectively.
Their amplitude and frequency regimes suggest that these are candidate blue large amplitude pulsators (BLAPs; \citealt{Pietrukowicz2017,Macfarlane2015}).

\subsection{Cataclysmic variables}\label{sec:cv}
cluster 2 consists of 296 objects, 140 of which are known CVs \citep{Barlow2022,Hou2023,Canbay2023} and 4 are candidate CVs from Krzesinski et al. (in prep). The remaining 152 objects are identified by SIMBAD as candidate hot subdwarfs (70), stars (61), variables (9), and CV candidates (3). We considered all of these objects as candidate CVs since all known objects in cluster 2 are CVs without contamination from other classes. The full lists of confirmed and candidate CVs are given in Table A.8 and A.9, respectively. 

By cross-matching the objects in cluster 2 with TESS, we found 127/140 confirmed CVs and 75/152 candidate CVs with TESS short-cadence light curves. Their period distributions are shown in Fig. \ref{fig:cvs_period_dist}, where the periods are centred at 3.43 hr and 4.63 hr for the known and candidate CVs, respectively. The 127/140 CVs represent the same objects as in the \citep{Canbay2023} catalogue. However, their reported periods are only available for 71 objects, mainly taken from \cite{Ritter2003}, with a median period of 3.40 hr. This means that we added 56 new candidate orbital periods from our analysis.
\begin{table*}
    \centering
    \caption{Variability classifications for known and candidate hot subdwarfs.}
    \label{tab:class_summary}
    \begin{tabular}{lccc|cccc|c}
\multicolumn{9}{c}{141/290 variable hot subdwarf candidates with {\it Gaia} and TESS lightcurves in cluster 0} \\ \hline
\multicolumn{4}{c|}{63 confirmed hot subdwarfs}&\multicolumn{4}{c|}{ 78 hot subdwarf candidates} \\ \hline

\multicolumn{3}{c}{Confirmed Variables}&\multicolumn{1}{c|}{ New Variables}&\multicolumn{2}{c}{Confirmed Variables}&\multicolumn{2}{c|}{ New Variables}&\multicolumn{1}{c}{Total new}  \\ \hline
Reflection &     15  &       &17 &  & 2&  &40 & 57 \\
HW Vir &     13  &       &6 &  & 7&  &14  &20\\
Ellipsoidal &     1  &       &5 &  & $-$&  &2&7  \\
Pulsating variables&     6  &       &$-$ &  & 3&  &$1$ & $1$ \\
Others/Unclear&     1 &       &1 &  & 5&  &5  &$-$\\
\hline\\

\multicolumn{9}{c}{149/290 variable hot subdwarf candidates with only {\it Gaia} lightcurves in cluster 0} \\ \hline
\multicolumn{1}{c}{}&\multicolumn{3}{c|}{15 confirmed hot subdwarfs}&\multicolumn{4}{c|}{134 hot subdwarf candidates}&\multicolumn{1}{c}{Total new} \\ \hline 
Sinusoidal &       &   5 & &  &  60& & & 65\\
HW Vir &       &     2 & &  &  20& & &22\\
Eclipsing binary &       &    5& &  & 14& & &19\\
Pulsating variables&       &   $-$ & &  & 2 & & &2\\
Others/Unclear&       &      3 & &  & 38& & &$-$\\
\hline\\

\end{tabular}
\end{table*}
\begin{figure}
  \centering
\includegraphics[width=0.95\linewidth]{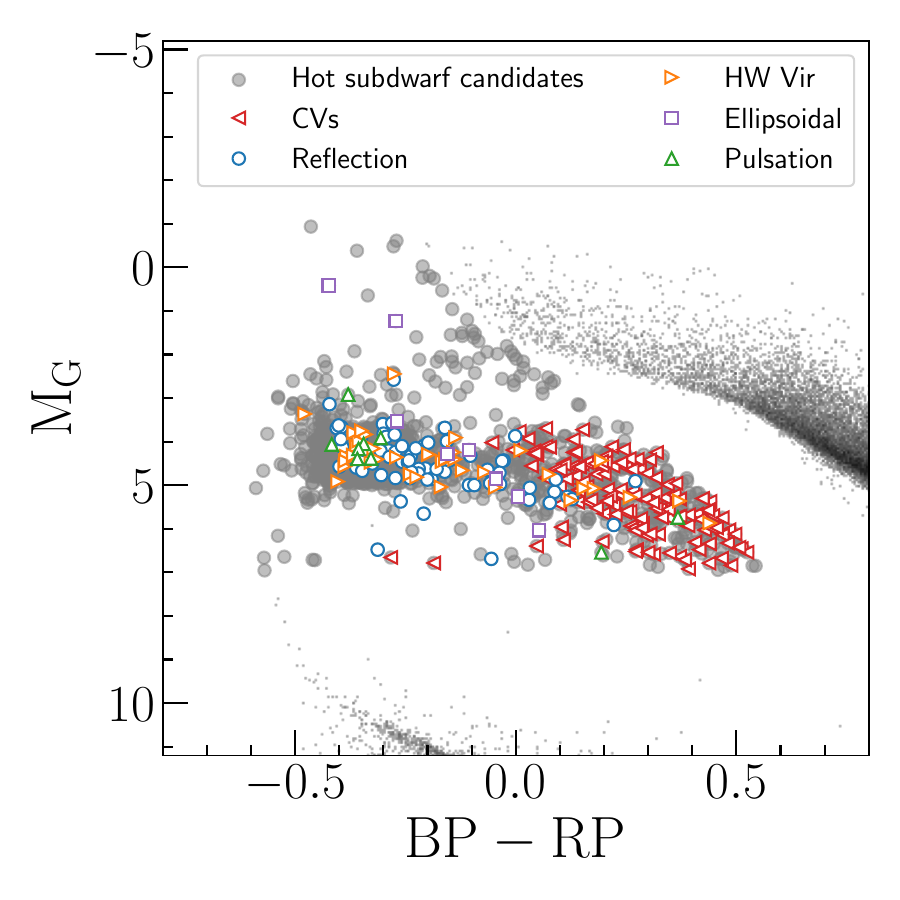}
\caption{\small {\it Gaia} DR3 colour-magnitude diagram depicting the candidate hot subdwarfs (1682) from \cite{Culpan2022} with {\it Gaia} light curves (grey circles). The variability classifications are shown for the selected candidate variable hot subdwarfs (290) with TESS observations (141/290). Among the candidate hot subdwarfs, CVs are also identified from the literature and are represented by left triangles. The grey background data points correspond to the {\it Gaia} Catalogue of Nearby Stars \citep{GaiaCollaboration2021}.
}\label{fig:cmd_culpan}
\end{figure}
\begin{figure}
  \centering
\includegraphics[width=0.95\linewidth]{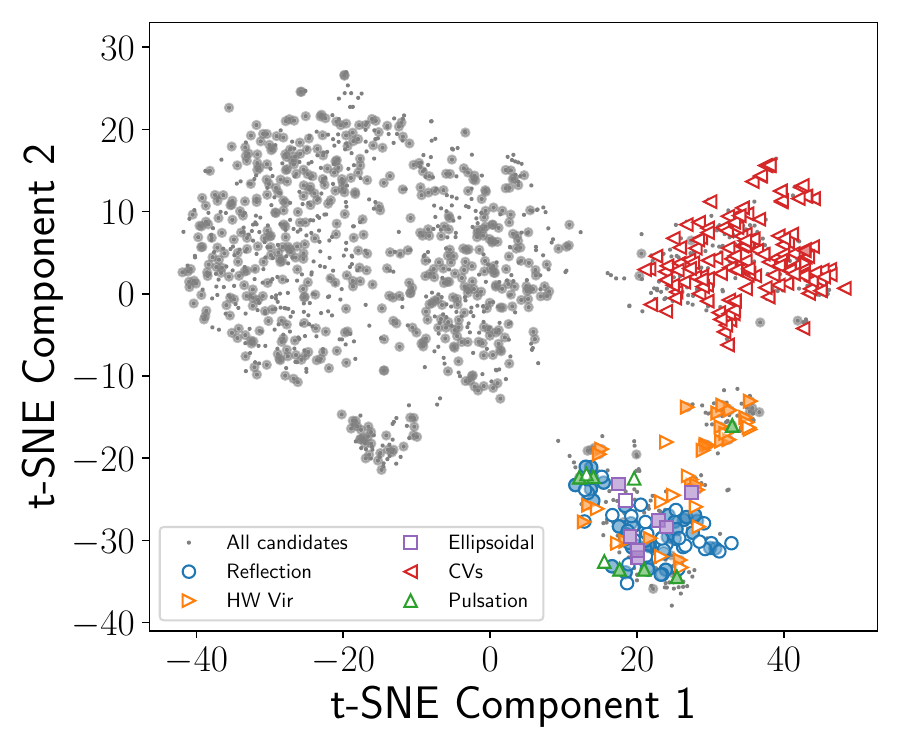}
\caption{\small Identified variables from {\it Gaia} and TESS light curves. The shaded colours correspond to confirmed hot subdwarfs. CV objects were obtained from the literature (see Sect. \ref{sec:cv}).}
\label{fig:cls_var_type}
\end{figure}
\begin{figure}
    \centering
    \includegraphics[width=0.95\linewidth]{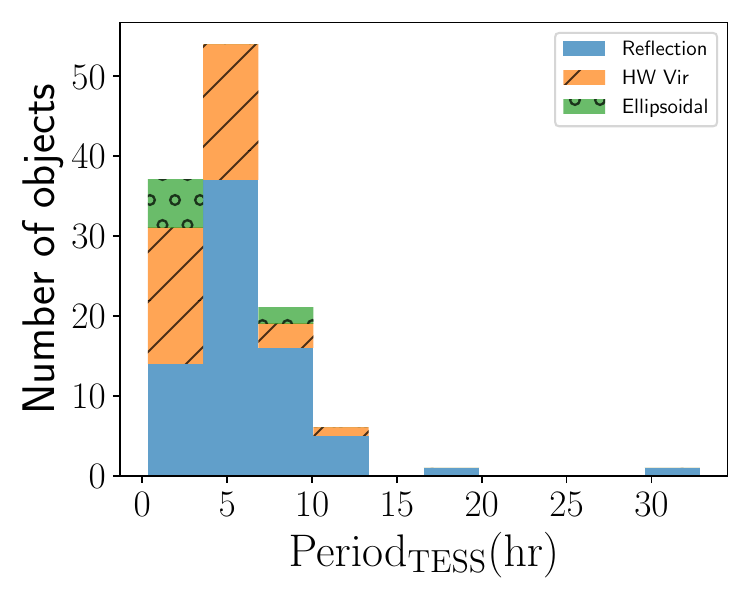}
    \caption{\small Period distribution of the binary systems observed with {\it Gaia} and TESS in cluster 0.}
    \label{fig:var_period_dist}
\end{figure}
\subsection{Variability distributions}\label{sec:discussion}
We investigated the photometric variability of 290 and 296 objects in cluster 0 and cluster 2, respectively. A summary of the variability classification of confirmed and candidate hot subdwarfs is presented in Table \ref{tab:class_summary}. In Fig. \ref{fig:cmd_culpan} we present a {\it Gaia} colour-magnitude diagram of the 1,576 candidate hot subdwarf variables (grey circles) with the {\it Gaia} Catalogue of Nearby Stars in the background (grey data points; \citealt{GaiaCollaboration2021}). Classified variables from cluster 0 with TESS light curves are shown in the figure. The light-curve shapes of reflection-effect systems can be explained by the fact that the hot subdwarf irradiates and heats one side of its cooler companion star, causing the cooler star to appear brighter on the side facing the hot subdwarf. As the system orbits, this creates a quasi-sinusoidal variability in the light curves. Depending on the viewing angle, reflection-effect systems can be eclipsing and form the HW Vir systems. On the other hand,  compact hot subdwarf binaries, particularly those with white dwarf companions, show ellipsoidal modulation in their light curves due to tidal distortion of the hot subdwarf, resulting in two maxima or two minima in their light curves. Examples of a reflection, HW Vir, and ellipsoidal system are shown in Fig.\ref{fig:sample_var}. As previously introduced, the evolutionary stages of these systems can be understood through the lens of a binary evolution channel, notably a common-envelope evolution for short-period systems. However, the exact formation mechanisms and evolutionary pathways are still areas of active research. On the other hand, CVs consist of a white dwarf primary and a mass-transferring secondary, typically a MS star. The shape of their light curves can mostly be explained by dramatic brightness increases known as outbursts, which are a result of instabilities in the accretion disk and lead to sudden higher mass transfer.
In Fig. \ref{fig:cmd_culpan}, reflection-effect and HW Vir systems appear to occupy the same area (centred at $\rm M_G=4.4$ and $\rm BP-RP=-0.2$) and tend to be bluer than the known CVs (centered at $\rm M_G=5.3$ and $\rm BP-RP=0.3$). 

Based on their locations in the t-SNE components, HW Vir systems tend to be more concentrated in the sub-cluster between cluster 0 and cluster 2, as shown in Fig. \ref{fig:cls_var_type}, with a broader G-magnitude range (\texttt{range\_mag\_g\_fov} around 0.50 mag) compared to the rest of the variables in cluster 0 (\texttt{range\_mag\_g\_fov} around 0.16 mag). Poor \textit{Gaia} sampling of HW Vir systems could result in a sinusoidal-like shape of their light curves, as shown in the first panel of Fig. 5, due to the smearing effect. This could place them in a different position in cluster 0 rather than in the sub-cluster. However, some HW Vir systems have shallower eclipse depths compared to others, and this could also place them in the main cluster in cluster 0. As previously mentioned, CVs lie in cluster 2 with a G-magnitude range, \texttt{range\_mag\_g\_fov}, centred at 1.15 mag. The distributions of the other features are presented in Fig. \ref{fig:features_median_iqr}, with the 10th percentile, the median, and the 90th percentile of the features for each cluster. In comparison to the other two clusters, cluster 2 exhibits a broader distribution of features, notably a high amplitude of variability, as shown in the right panel of Fig.\ref{fig:cls_amp_gradient} and Fig. \ref{fig:features_median_iqr}. These differences in the feature distributions could be relevant for the reduction algorithms to represent the clusters in the low-dimensional space well.

Of the new variables identified from {\it Gaia} and TESS in cluster 0, $\sim 23 \%$ are classified as HW Vir systems, $\sim 67\%$ are reflection-effect systems, and $\sim 10\%$ are ellipsoidal and pulsating variables. For their period distributions, Fig. \ref{fig:var_period_dist} shows that the periods of known and new HW Vir systems are in the range of $\sim 1.5$ hr to $\sim 9$ hr, while those of the reflection-effect systems range from $\sim 1.7$ hr to $\sim 35$ hr. This difference in the period distribution of the eclipsing reflection-effect (HW Vir) and non-eclipsing reflection-effect systems has also been observed in other studies. HW Vir systems tend to have shorter periods than non-eclipsing reflection-effect systems, as found by \citep{Schaffenroth2022}. These authors also found a broad peak at periods from 2 to 8 hr, but were unable to find objects with a period longer than $\sim 30$ hr for reflection-effect systems. They reported that periods longer than a few days might be rare or might not exist for these systems. However, we found several objects with periods longer than a few days from {\it Gaia}, which might be binary or eclipsing systems. Since we have no TESS observations for these objects, their variability types are referred to as sinusoidal or eclipsing binary.
 
\section{Conclusion and future prospects}\label{sec:conclusion}
We set out to develop a machine-learning algorithm that might be generalised and that leverages multi-band photometric time-series data in order to classify variable and non-variable subdwarfs. We developed our algorithm using multi-band time-series photometry from {\it Gaia} and validated the algorithm using independent {\it TESS} data. Starting with a readily available catalogue of 61,585 candidate hot subdwarfs, we were able to extract {\it Gaia} multi-band light curves of 1,682 objects with good astrometric solutions and a variable number of observations in the {\it Gaia} photometric bands (with 25 observations at least). We searched for periodicities using the hybrid $\Psi-$statistic approach and estimated the uncertainties associated with the determined frequencies with a Monte Carlo approach. 

Using the sparsely sampled multi-band {\it Gaia} photometric data, we defined a number of bespoke summary statistics to supplement those already provided by the {\it Gaia} database. We applied machine-learning algorithms to calculate the importance of the feature and reduce the dimensionality before we applied a  clustering algorithm that identified three clusters, which are predominantly predicted by the amplitude of the photometric variability in the {\it Gaia} G band. We further validated the results by applying two different dimensionality reduction techniques, which resulted in 99\% similar results.

The three clusters that we identified correspond to (candidate) hot subdwarfs with statistically significant variability (cluster 0), non-variable subwarfs (cluster 1), and CVs (cluster 2). 

Upon further inspection, we were able to identify different populations of variable hot subdwarfs observed from {\it Gaia} and TESS in cluster 0. A significant number of them are in binaries, while a few pulsating variables are detected. The scarcity of the observed pulsating variables in {\it Gaia} could be explained by the fact that hot subdwarfs pulsate with low-amplitude light variations  of about a few milli-magnitudes.

Our analysis allowed us to newly identify a large number stars as variables, notably reflection-effect and HW Vir systems. The key findings of the clustering analysis are summarised below.
\begin{itemize}
    \item In cluster 0, 89 new hot subdwarf variables were identified from {\it Gaia} and TESS observations, while 108 new variables were found from {\it Gaia} alone. These new variables are mainly reflection-effect and HW Vir systems.
    \item In the same cluster 0, nine previously identified pulsating variables were found among the candidate variable hot subdwarf. We further identified two new high-amplitude pulsating objects that are consistent with being BLAPs.
    \item In cluster 2, a large number of CVs were identified, of which 140 were spectroscopically confirmed in other studies. We consider the remaining 156 objects in cluster 2 to be candidate CVs.
    \item Feature evaluation based on the three clusters showed that features related to the photometric variations in the G band strongly contribute to characterising the clusters, including the amplitude, the magnitude range, and the interquartile range of the G-band light curves. The G-band amplitude distribution suggests a lower limit of $\sim 0.02$ mag on the detection of clear variability in the light curve.
\end{itemize}

The classification algorithm developed in this work was specifically designed to be flexible and generalisable. We used widely available features and developed new features that can be efficiently calculated for independent data sets with different properties. As a result of this, we can include new observations and objects without having to retrain the algorithm. Furthermore, our results can be used to help build labelled datasets for future supervised machine-learning classifications of variable stars.

Scientifically, our results are twofold. First, we developed a robust method for identifying variable subdwarf stars. Second, we developed an algorithm that efficiently identifies CVs without the need for expensive follow-up spectroscopic observations. Together, these results allowed us to confidently identify new variable subdwarfs for further analysis from existing data while filtering out contaminating sources such as CVs. While hundreds of hot subdwarfs and CVs have already been discovered, a systematically discovered sample of these objects is required to better understand various binary interaction processes, such as mass transfer, common-envelope evolution, and tidal interactions. Furthermore, an algorithm that efficiently identifies variable and non-variable subdwarfs from sparsely sampled data with known amplitude biases offers a unique opportunity for building observational instability strips. By increasing the number of known sdBVs, we can perform population-level asteroseismic studies, similar to the work done for $\beta$ Cep stars using {\it Gaia} and TESS data \citep{Fritzewski2024}. This approach has the potential to reveal new insights into the pulsation properties and interior structure of hot subdwarfs by leveraging multi-colour photometry and observational amplitude ratios for mode identifications.

Spectroscopic follow-up observations, such as those with the 4-metre Multi-Object Spectroscopic Telescope (4MOST;\citealt{deJong2019}), the William Herschel Telescope Enhanced Area Velocity Explorer (WEAVE; \citealt{Jin2024}), the Sloan Digital Sky Survey V (SDSS-V; \citealt{Kollmeier2019}), and the Large sky Area Multi-Object fiber Spectroscopic Telescope (LAMOST; \citealt{Cui2012}) may deliver radial velocity data and atmospheric parameters to confirm the physical nature of these new variables (153 candidate hot subdwarf and 152 candidate CVs), as well as the two new high-amplitude pulsating variables identified from {\it Gaia}. Other future prospects include photometric observations of the pulsating variables identified in this work using BlackGEM \citep{Groot2024} to obtain multi-band pulsation amplitudes for mode identifications and asteroseismic modelling. Additionally, the release of {\it Gaia} Data Release 4 (DR4), which will include all photometric data, offers a valuable prospect for further exploration. When the complete photometric dataset becomes available, this work can immediately be applied to the remaining 59\,471 objects, enabling a comprehensive analysis of variability across a wider range of sources.

\section{Data availability}\label{sec:data_availability}
Tables A.4 to A.10 are only available in electronic form at the CDS via anonymous ftp to cdsarc.u-strasbg.fr (130.79.128.5) or via \url{https://cdsarc.cds.unistra.fr/viz-bin/cat/J/A+A/693/A268}.

\begin{acknowledgements}
TK acknowledges support from the National Science Foundation through grant AST \#2107982, from NASA through grant 80NSSC22K0338 and from STScI through grant HST-GO-16659.002-A. Co-funded by the European Union (ERC, CompactBINARIES, 101078773). Views and opinions expressed are however those of the author(s) only and do not necessarily reflect those of the European Union or the European Research Council. Neither the European Union nor the granting authority can be held responsible for them. The research leading to these results has received funding from the Research Foundation Flanders (FWO) under grant agreement
G0A2917N (BlackGEM), as well as from the BELgian federal Science Policy Office (BELSPO) through PRODEX grants for {\it Gaia} data exploitation.
This work has made use of data from the European Space Agency (ESA) mission {\it Gaia} (https://www.cosmos.esa.int/Gaia), processed by the {\it Gaia} Data Processing and Analysis Consortium (DPAC, https://www.cosmos.esa.int/web/Gaia/dpac/consortium). Funding for the DPAC has been provided by national institutions, in particular the institutions participating in the {\it Gaia} Multilateral Agreement. PJG is supported by NRF SARChI grant 111692. 

\end{acknowledgements}

\bibliographystyle{aa}
\bibliography{aanda.bib}

\begin{thebibliography}{87}
\expandafter\ifx\csname natexlab\endcsname\relax\def\natexlab#1{#1}\fi

\bibitem[{{Aerts} \& {Tkachenko}(2023)}]{Aerts2023}
{Aerts}, C. \& {Tkachenko}, A. 2023, arXiv e-prints, arXiv:2311.08453

\bibitem[{{Baluev}(2008)}]{Baluev2008}
{Baluev}, R.~V. 2008, \mnras, 385, 1279

\bibitem[{{Barlow} {et~al.}(2022){Barlow}, {Corcoran}, {Parker}, {Kupfer}, {N{\'e}meth}, {Hermes}, {Lopez}, {Frondorf}, {Vestal}, \& {Holden}}]{Barlow2022}
{Barlow}, B.~N., {Corcoran}, K.~A., {Parker}, I.~M., {et~al.} 2022, \apj, 928, 20

\bibitem[{{Bellm} {et~al.}(2019){Bellm}, {Kulkarni}, {Graham}, {Dekany}, {Smith}, {Riddle}, {Masci}, {Helou}, {Prince}, {Adams}, {Barbarino}, {Barlow}, {Bauer}, {Beck}, {Belicki}, {Biswas}, {Blagorodnova}, {Bodewits}, {Bolin}, {Brinnel}, {Brooke}, {Bue}, {Bulla}, {Burruss}, {Cenko}, {Chang}, {Connolly}, {Coughlin}, {Cromer}, {Cunningham}, {De}, {Delacroix}, {Desai}, {Duev}, {Eadie}, {Farnham}, {Feeney}, {Feindt}, {Flynn}, {Franckowiak}, {Frederick}, {Fremling}, {Gal-Yam}, {Gezari}, {Giomi}, {Goldstein}, {Golkhou}, {Goobar}, {Groom}, {Hacopians}, {Hale}, {Henning}, {Ho}, {Hover}, {Howell}, {Hung}, {Huppenkothen}, {Imel}, {Ip}, {Ivezi{\'c}}, {Jackson}, {Jones}, {Juric}, {Kasliwal}, {Kaspi}, {Kaye}, {Kelley}, {Kowalski}, {Kramer}, {Kupfer}, {Landry}, {Laher}, {Lee}, {Lin}, {Lin}, {Lunnan}, {Giomi}, {Mahabal}, {Mao}, {Miller}, {Monkewitz}, {Murphy}, {Ngeow}, {Nordin}, {Nugent}, {Ofek}, {Patterson}, {Penprase}, {Porter}, {Rauch}, {Rebbapragada}, {Reiley}, {Rigault}, {Rodriguez}, {van Roestel}, {Rusholme}, {van
  Santen}, {Schulze}, {Shupe}, {Singer}, {Soumagnac}, {Stein}, {Surace}, {Sollerman}, {Szkody}, {Taddia}, {Terek}, {Van Sistine}, {van Velzen}, {Vestrand}, {Walters}, {Ward}, {Ye}, {Yu}, {Yan}, \& {Zolkower}}]{Bellm2019}
{Bellm}, E.~C., {Kulkarni}, S.~R., {Graham}, M.~J., {et~al.} 2019, \pasp, 131, 018002

\bibitem[{{Bloemen} {et~al.}(2014){Bloemen}, {Hu}, {Aerts}, {Dupret}, {{\O}stensen}, {Degroote}, {M{\"u}ller-Ringat}, \& {Rauch}}]{Bloemen2014}
{Bloemen}, S., {Hu}, H., {Aerts}, C., {et~al.} 2014, \aap, 569, A123

\bibitem[{{Brassard} {et~al.}(2001){Brassard}, {Fontaine}, {Bill{\`e}res}, {Charpinet}, {Liebert}, \& {Saffer}}]{Brassard2001}
{Brassard}, P., {Fontaine}, G., {Bill{\`e}res}, M., {et~al.} 2001, \apj, 563, 1013

\bibitem[{Breiman(2001)}]{breiman2001}
Breiman, L. 2001, Machine Learning, 45, 5

\bibitem[{{Canbay} {et~al.}(2023){Canbay}, {Bilir}, {{\"O}zd{\"o}nmez}, \& {Ak}}]{Canbay2023}
{Canbay}, R., {Bilir}, S., {{\"O}zd{\"o}nmez}, A., \& {Ak}, T. 2023, \aj, 165, 163

\bibitem[{{Charpinet} {et~al.}(1997){Charpinet}, {Fontaine}, {Brassard}, {Chayer}, {Rogers}, {Iglesias}, \& {Dorman}}]{Charpinet1997}
{Charpinet}, S., {Fontaine}, G., {Brassard}, P., {et~al.} 1997, \apjl, 483, L123

\bibitem[{{Charpinet} {et~al.}(2010){Charpinet}, {Green}, {Baglin}, {Van Grootel}, {Fontaine}, {Vauclair}, {Chaintreuil}, {Weiss}, {Michel}, {Auvergne}, {Catala}, {Samadi}, \& {Baudin}}]{Charpinet2010}
{Charpinet}, S., {Green}, E.~M., {Baglin}, A., {et~al.} 2010, \aap, 516, L6

\bibitem[{{Cui} {et~al.}(2022){Cui}, {Liu}, {Feng}, \& {Liu}}]{Cui2022}
{Cui}, K., {Liu}, J., {Feng}, F., \& {Liu}, J. 2022, \aj, 163, 23

\bibitem[{{Cui} {et~al.}(2012){Cui}, {Zhao}, {Chu}, {Li}, {Li}, {Zhang}, {Su}, {Yao}, {Wang}, {Xing}, {Li}, {Zhu}, {Wang}, {Gu}, {Luo}, {Xu}, {Zhang}, {Liu}, {Zhang}, {Yang}, {Cao}, {Chen}, {Chen}, {Chen}, {Chen}, {Chu}, {Feng}, {Gong}, {Hou}, {Hu}, {Hu}, {Hu}, {Jia}, {Jiang}, {Jiang}, {Jiang}, {Jin}, {Li}, {Li}, {Li}, {Liu}, {Liu}, {Lu}, {Mao}, {Men}, {Qi}, {Qi}, {Shi}, {Tang}, {Tao}, {Wang}, {Wang}, {Wang}, {Wang}, {Wang}, {Wang}, {Wang}, {Wang}, {Wang}, {Wang}, {Wang}, {Wang}, {Xu}, {Xu}, {Yang}, {Yu}, {Yuan}, {Yuan}, {Zhai}, {Zhang}, {Zhang}, {Zhang}, {Zhao}, {Zhou}, {Zhou}, {Zhu}, \& {Zou}}]{Cui2012}
{Cui}, X.-Q., {Zhao}, Y.-H., {Chu}, Y.-Q., {et~al.} 2012, Research in Astronomy and Astrophysics, 12, 1197

\bibitem[{{Culpan} {et~al.}(2022){Culpan}, {Geier}, {Reindl}, {Pelisoli}, {Gentile Fusillo}, \& {Vorontseva}}]{Culpan2022}
{Culpan}, R., {Geier}, S., {Reindl}, N., {et~al.} 2022, \aap, 662, A40

\bibitem[{{Dawson} {et~al.}(2024){Dawson}, {Geier}, {Heber}, {Pelisoli}, {Dorsch}, {Schaffenroth}, {Reindl}, {Culpan}, {Pritzkuleit}, {Vos}, {Soemitro}, {Roth}, {Schneider}, {Uzundag}, {Vu{\v{c}}kovi{\'c}}, {Antunes Amaral}, {Istrate}, {Justham}, {{\O}stensen}, {Telting}, {Djupvik}, {Raddi}, {Green}, {Jeffery}, {Kepler}, {Munday}, {Steinmetz}, \& {Kupfer}}]{Dawson2024}
{Dawson}, H., {Geier}, S., {Heber}, U., {et~al.} 2024, \aap, 686, A25

\bibitem[{{de Jong} {et~al.}(2019){de Jong}, {Agertz}, {Berbel}, {Aird}, {Alexander}, {Amarsi}, {Anders}, {Andrae}, {Ansarinejad}, {Ansorge}, {Antilogus}, {Anwand-Heerwart}, {Arentsen}, {Arnadottir}, {Asplund}, {Auger}, {Azais}, {Baade}, {Baker}, {Baker}, {Balbinot}, {Baldry}, {Banerji}, {Barden}, {Barklem}, {Barth{\'e}l{\'e}my-Mazot}, {Battistini}, {Bauer}, {Bell}, {Bellido-Tirado}, {Bellstedt}, {Belokurov}, {Bensby}, {Bergemann}, {Bestenlehner}, {Bielby}, {Bilicki}, {Blake}, {Bland-Hawthorn}, {Boeche}, {Boland}, {Boller}, {Bongard}, {Bongiorno}, {Bonifacio}, {Boudon}, {Brooks}, {Brown}, {Brown}, {Br{\"u}ggen}, {Brynnel}, {Brzeski}, {Buchert}, {Buschkamp}, {Caffau}, {Caillier}, {Carrick}, {Casagrande}, {Case}, {Casey}, {Cesarini}, {Cescutti}, {Chapuis}, {Chiappini}, {Childress}, {Christlieb}, {Church}, {Cioni}, {Cluver}, {Colless}, {Collett}, {Comparat}, {Cooper}, {Couch}, {Courbin}, {Croom}, {Croton}, {Daguis{\'e}}, {Dalton}, {Davies}, {Davis}, {de Laverny}, {Deason}, {Dionies}, {Disseau}, {Doel},
  {D{\"o}scher}, {Driver}, {Dwelly}, {Eckert}, {Edge}, {Edvardsson}, {Youssoufi}, {Elhaddad}, {Enke}, {Erfanianfar}, {Farrell}, {Fechner}, {Feiz}, {Feltzing}, {Ferreras}, {Feuerstein}, {Feuillet}, {Finoguenov}, {Ford}, {Fotopoulou}, {Fouesneau}, {Frenk}, {Frey}, {Gaessler}, {Geier}, {Gentile Fusillo}, {Gerhard}, {Giannantonio}, {Giannone}, {Gibson}, {Gillingham}, {Gonz{\'a}lez-Fern{\'a}ndez}, {Gonzalez-Solares}, {Gottloeber}, {Gould}, {Grebel}, {Gueguen}, {Guiglion}, {Haehnelt}, {Hahn}, {Hansen}, {Hartman}, {Hauptner}, {Hawkins}, {Haynes}, {Haynes}, {Heiter}, {Helmi}, {Aguayo}, {Hewett}, {Hinton}, {Hobbs}, {Hoenig}, {Hofman}, {Hook}, {Hopgood}, {Hopkins}, {Hourihane}, {Howes}, {Howlett}, {Huet}, {Irwin}, {Iwert}, {Jablonka}, {Jahn}, {Jahnke}, {Jarno}, {Jin}, {Jofre}, {Johl}, {Jones}, {J{\"o}nsson}, {Jordan}, {Karovicova}, {Khalatyan}, {Kelz}, {Kennicutt}, {King}, {Kitaura}, {Klar}, {Klauser}, {Kneib}, {Koch}, {Koposov}, {Kordopatis}, {Korn}, {Kosmalski}, {Kotak}, {Kovalev}, {Kreckel}, {Kripak}, {Krumpe},
  {Kuijken}, {Kunder}, {Kushniruk}, {Lam}, {Lamer}, {Laurent}, {Lawrence}, {Lehmitz}, {Lemasle}, {Lewis}, {Li}, {Lidman}, {Lind}, {Liske}, {Lizon}, {Loveday}, {Ludwig}, {McDermid}, {Maguire}, {Mainieri}, {Mali}, {Mandel}, {Mandel}, {Mannering}, {Martell}, {Martinez Delgado}, {Matijevic}, {McGregor}, {McMahon}, {McMillan}, {Mena}, {Merloni}, {Meyer}, {Michel}, {Micheva}, {Migniau}, {Minchev}, {Monari}, {Muller}, {Murphy}, {Muthukrishna}, {Nandra}, {Navarro}, {Ness}, {Nichani}, {Nichol}, {Nicklas}, {Niederhofer}, {Norberg}, {Obreschkow}, {Oliver}, {Owers}, {Pai}, {Pankratow}, {Parkinson}, {Paschke}, {Paterson}, {Pecontal}, {Parry}, {Phillips}, {Pillepich}, {Pinard}, {Pirard}, {Piskunov}, {Plank}, {Pl{\"u}schke}, {Pons}, {Popesso}, {Power}, {Pragt}, {Pramskiy}, {Pryer}, {Quattri}, {Queiroz}, {Quirrenbach}, {Rahurkar}, {Raichoor}, {Ramstedt}, {Rau}, {Recio-Blanco}, {Reiss}, {Renaud}, {Revaz}, {Rhode}, {Richard}, {Richter}, {Rix}, {Robotham}, {Roelfsema}, {Romaniello}, {Rosario}, {Rothmaier}, {Roukema}, {Ruchti},
  {Rupprecht}, {Rybizki}, {Ryde}, {Saar}, {Sadler}, {Sahl{\'e}n}, {Salvato}, {Sassolas}, {Saunders}, {Saviauk}, {Sbordone}, {Schmidt}, {Schnurr}, {Scholz}, {Schwope}, {Seifert}, {Shanks}, {Sheinis}, {Sivov}, {Sk{\'u}lad{\'o}ttir}, {Smartt}, {Smedley}, {Smith}, {Smith}, {Sorce}, {Spitler}, {Starkenburg}, {Steinmetz}, {Stilz}, {Storm}, {Sullivan}, {Sutherland}, {Swann}, {Tamone}, {Taylor}, {Teillon}, {Tempel}, {ter Horst}, {Thi}, {Tolstoy}, {Trager}, {Traven}, {Tremblay}, {Tresse}, {Valentini}, {van de Weygaert}, {van den Ancker}, {Veljanoski}, {Venkatesan}, {Wagner}, {Wagner}, {Walcher}, {Waller}, {Walton}, {Wang}, {Winkler}, {Wisotzki}, {Worley}, {Worseck}, {Xiang}, {Xu}, {Yong}, {Zhao}, {Zheng}, {Zscheyge}, \& {Zucker}}]{deJong2019}
{de Jong}, R.~S., {Agertz}, O., {Berbel}, A.~A., {et~al.} 2019, The Messenger, 175, 3

\bibitem[{{Deca} {et~al.}(2012){Deca}, {Marsh}, {{\O}stensen}, {Morales-Rueda}, {Copperwheat}, {Wade}, {Stark}, {Maxted}, {Nelemans}, \& {Heber}}]{Deca2012}
{Deca}, J., {Marsh}, T.~R., {{\O}stensen}, R.~H., {et~al.} 2012, \mnras, 421, 2798

\bibitem[{{Dorman} {et~al.}(1993){Dorman}, {Rood}, \& {O'Connell}}]{Dorman1993}
{Dorman}, B., {Rood}, R.~T., \& {O'Connell}, R.~W. 1993, \apj, 419, 596

\bibitem[{{Eyer} {et~al.}(2023){Eyer}, {Audard}, {Holl}, {Rimoldini}, {Carnerero}, {Clementini}, {De Ridder}, {Distefano}, {Evans}, {Gavras}, {Gomel}, {Lebzelter}, {Marton}, {Mowlavi}, {Panahi}, {Ripepi}, {Wyrzykowski}, {Nienartowicz}, {Jevardat de Fombelle}, {Lecoeur-Taibi}, {Rohrbasser}, {Riello}, {Garc{\'\i}a-Lario}, {Lanzafame}, {Mazeh}, {Raiteri}, {Zucker}, {{\'A}brah{\'a}m}, {Aerts}, {Aguado}, {Anderson}, {Bashi}, {Binnenfeld}, {Faigler}, {Garofalo}, {Karbevska}, {K{\'o}sp{\'a}l}, {Kruszy{\'n}ska}, {Kun}, {Lanza}, {Leccia}, {Marconi}, {Messina}, {Molinaro}, {Moln{\'a}r}, {Muraveva}, {Musella}, {Nagy}, {Pagano}, {Palaversa}, {Plachy}, {Pr{\v{s}}a}, {Rybicki}, {Shahaf}, {Szabados}, {Szegedi-Elek}, {Trabucchi}, {Barblan}, {Grenon}, {Roelens}, \& {S{\"u}veges}}]{Eyer2023}
{Eyer}, L., {Audard}, M., {Holl}, B., {et~al.} 2023, \aap, 674, A13

\bibitem[{{Fontaine} {et~al.}(2003){Fontaine}, {Brassard}, {Charpinet}, {Green}, {Chayer}, {Bill{\`e}res}, \& {Randall}}]{Fontaine2003}
{Fontaine}, G., {Brassard}, P., {Charpinet}, S., {et~al.} 2003, \apj, 597, 518

\bibitem[{{Fritzewski} {et~al.}(2024){Fritzewski}, {Vanrespaille}, {Aerts}, {Hey}, \& {De Ridder}}]{Fritzewski2024}
{Fritzewski}, D.~J., {Vanrespaille}, M., {Aerts}, C., {Hey}, D., \& {De Ridder}, J. 2024, arXiv e-prints, arXiv:2408.06097

\bibitem[{{Gaia Collaboration} {et~al.}(2018){Gaia Collaboration}, {Brown}, {Vallenari}, {Prusti}, {de Bruijne}, {Babusiaux}, {Bailer-Jones}, {Biermann}, {Evans}, {Eyer}, {Jansen}, {Jordi}, {Klioner}, {Lammers}, {Lindegren}, {Luri}, {Mignard}, {Panem}, {Pourbaix}, {Randich}, {Sartoretti}, {Siddiqui}, {Soubiran}, {van Leeuwen}, {Walton}, {Arenou}, {Bastian}, {Cropper}, {Drimmel}, {Katz}, {Lattanzi}, {Bakker}, {Cacciari}, {Casta{\~n}eda}, {Chaoul}, {Cheek}, {De Angeli}, {Fabricius}, {Guerra}, {Holl}, {Masana}, {Messineo}, {Mowlavi}, {Nienartowicz}, {Panuzzo}, {Portell}, {Riello}, {Seabroke}, {Tanga}, {Th{\'e}venin}, {Gracia-Abril}, {Comoretto}, {Garcia-Reinaldos}, {Teyssier}, {Altmann}, {Andrae}, {Audard}, {Bellas-Velidis}, {Benson}, {Berthier}, {Blomme}, {Burgess}, {Busso}, {Carry}, {Cellino}, {Clementini}, {Clotet}, {Creevey}, {Davidson}, {De Ridder}, {Delchambre}, {Dell'Oro}, {Ducourant}, {Fern{\'a}ndez-Hern{\'a}ndez}, {Fouesneau}, {Fr{\'e}mat}, {Galluccio}, {Garc{\'\i}a-Torres},
  {Gonz{\'a}lez-N{\'u}{\~n}ez}, {Gonz{\'a}lez-Vidal}, {Gosset}, {Guy}, {Halbwachs}, {Hambly}, {Harrison}, {Hern{\'a}ndez}, {Hestroffer}, {Hodgkin}, {Hutton}, {Jasniewicz}, {Jean-Antoine-Piccolo}, {Jordan}, {Korn}, {Krone-Martins}, {Lanzafame}, {Lebzelter}, {L{\"o}ffler}, {Manteiga}, {Marrese}, {Mart{\'\i}n-Fleitas}, {Moitinho}, {Mora}, {Muinonen}, {Osinde}, {Pancino}, {Pauwels}, {Petit}, {Recio-Blanco}, {Richards}, {Rimoldini}, {Robin}, {Sarro}, {Siopis}, {Smith}, {Sozzetti}, {S{\"u}veges}, {Torra}, {van Reeven}, {Abbas}, {Abreu Aramburu}, {Accart}, {Aerts}, {Altavilla}, {{\'A}lvarez}, {Alvarez}, {Alves}, {Anderson}, {Andrei}, {Anglada Varela}, {Antiche}, {Antoja}, {Arcay}, {Astraatmadja}, {Bach}, {Baker}, {Balaguer-N{\'u}{\~n}ez}, {Balm}, {Barache}, {Barata}, {Barbato}, {Barblan}, {Barklem}, {Barrado}, {Barros}, {Barstow}, {Bartholom{\'e} Mu{\~n}oz}, {Bassilana}, {Becciani}, {Bellazzini}, {Berihuete}, {Bertone}, {Bianchi}, {Bienaym{\'e}}, {Blanco-Cuaresma}, {Boch}, {Boeche}, {Bombrun}, {Borrachero},
  {Bossini}, {Bouquillon}, {Bourda}, {Bragaglia}, {Bramante}, {Breddels}, {Bressan}, {Brouillet}, {Br{\"u}semeister}, {Brugaletta}, {Bucciarelli}, {Burlacu}, {Busonero}, {Butkevich}, {Buzzi}, {Caffau}, {Cancelliere}, {Cannizzaro}, {Cantat-Gaudin}, {Carballo}, {Carlucci}, {Carrasco}, {Casamiquela}, {Castellani}, {Castro-Ginard}, {Charlot}, {Chemin}, {Chiavassa}, {Cocozza}, {Costigan}, {Cowell}, {Crifo}, {Crosta}, {Crowley}, {Cuypers}, {Dafonte}, {Damerdji}, {Dapergolas}, {David}, {David}, {de Laverny}, {De Luise}, {De March}, {de Martino}, {de Souza}, {de Torres}, {Debosscher}, {del Pozo}, {Delbo}, {Delgado}, {Delgado}, {Di Matteo}, {Diakite}, {Diener}, {Distefano}, {Dolding}, {Drazinos}, {Dur{\'a}n}, {Edvardsson}, {Enke}, {Eriksson}, {Esquej}, {Eynard Bontemps}, {Fabre}, {Fabrizio}, {Faigler}, {Falc{\~a}o}, {Farr{\`a}s Casas}, {Federici}, {Fedorets}, {Fernique}, {Figueras}, {Filippi}, {Findeisen}, {Fonti}, {Fraile}, {Fraser}, {Fr{\'e}zouls}, {Gai}, {Galleti}, {Garabato}, {Garc{\'\i}a-Sedano}, {Garofalo},
  {Garralda}, {Gavel}, {Gavras}, {Gerssen}, {Geyer}, {Giacobbe}, {Gilmore}, {Girona}, {Giuffrida}, {Glass}, {Gomes}, {Granvik}, {Gueguen}, {Guerrier}, {Guiraud}, {Guti{\'e}rrez-S{\'a}nchez}, {Haigron}, {Hatzidimitriou}, {Hauser}, {Haywood}, {Heiter}, {Helmi}, {Heu}, {Hilger}, {Hobbs}, {Hofmann}, {Holland}, {Huckle}, {Hypki}, {Icardi}, {Jan{\ss}en}, {Jevardat de Fombelle}, {Jonker}, {Juh{\'a}sz}, {Julbe}, {Karampelas}, {Kewley}, {Klar}, {Kochoska}, {Kohley}, {Kolenberg}, {Kontizas}, {Kontizas}, {Koposov}, {Kordopatis}, {Kostrzewa-Rutkowska}, {Koubsky}, {Lambert}, {Lanza}, {Lasne}, {Lavigne}, {Le Fustec}, {Le Poncin-Lafitte}, {Lebreton}, {Leccia}, {Leclerc}, {Lecoeur-Taibi}, {Lenhardt}, {Leroux}, {Liao}, {Licata}, {Lindstr{\o}m}, {Lister}, {Livanou}, {Lobel}, {L{\'o}pez}, {Managau}, {Mann}, {Mantelet}, {Marchal}, {Marchant}, {Marconi}, {Marinoni}, {Marschalk{\'o}}, {Marshall}, {Martino}, {Marton}, {Mary}, {Massari}, {Matijevi{\v{c}}}, {Mazeh}, {McMillan}, {Messina}, {Michalik}, {Millar}, {Molina}, {Molinaro},
  {Moln{\'a}r}, {Montegriffo}, {Mor}, {Morbidelli}, {Morel}, {Morris}, {Mulone}, {Muraveva}, {Musella}, {Nelemans}, {Nicastro}, {Noval}, {O'Mullane}, {Ord{\'e}novic}, {Ord{\'o}{\~n}ez-Blanco}, {Osborne}, {Pagani}, {Pagano}, {Pailler}, {Palacin}, {Palaversa}, {Panahi}, {Pawlak}, {Piersimoni}, {Pineau}, {Plachy}, {Plum}, {Poggio}, {Poujoulet}, {Pr{\v{s}}a}, {Pulone}, {Racero}, {Ragaini}, {Rambaux}, {Ramos-Lerate}, {Regibo}, {Reyl{\'e}}, {Riclet}, {Ripepi}, {Riva}, {Rivard}, {Rixon}, {Roegiers}, {Roelens}, {Romero-G{\'o}mez}, {Rowell}, {Royer}, {Ruiz-Dern}, {Sadowski}, {Sagrist{\`a} Sell{\'e}s}, {Sahlmann}, {Salgado}, {Salguero}, {Sanna}, {Santana-Ros}, {Sarasso}, {Savietto}, {Schultheis}, {Sciacca}, {Segol}, {Segovia}, {S{\'e}gransan}, {Shih}, {Siltala}, {Silva}, {Smart}, {Smith}, {Solano}, {Solitro}, {Sordo}, {Soria Nieto}, {Souchay}, {Spagna}, {Spoto}, {Stampa}, {Steele}, {Steidelm{\"u}ller}, {Stephenson}, {Stoev}, {Suess}, {Surdej}, {Szabados}, {Szegedi-Elek}, {Tapiador}, {Taris}, {Tauran}, {Taylor},
  {Teixeira}, {Terrett}, {Teyssandier}, {Thuillot}, {Titarenko}, {Torra Clotet}, {Turon}, {Ulla}, {Utrilla}, {Uzzi}, {Vaillant}, {Valentini}, {Valette}, {van Elteren}, {Van Hemelryck}, {van Leeuwen}, {Vaschetto}, {Vecchiato}, {Veljanoski}, {Viala}, {Vicente}, {Vogt}, {von Essen}, {Voss}, {Votruba}, {Voutsinas}, {Walmsley}, {Weiler}, {Wertz}, {Wevers}, {Wyrzykowski}, {Yoldas}, {{\v{Z}}erjal}, {Ziaeepour}, {Zorec}, {Zschocke}, {Zucker}, {Zurbach}, \& {Zwitter}}]{GaiaCollab2018}
{Gaia Collaboration}, {Brown}, A.~G.~A., {Vallenari}, A., {et~al.} 2018, \aap, 616, A1

\bibitem[{{Gaia Collaboration} {et~al.}(2021{\natexlab{a}}){Gaia Collaboration}, {Brown}, {Vallenari}, {Prusti}, {de Bruijne}, {Babusiaux}, {Biermann}, {Creevey}, {Evans}, {Eyer}, {Hutton}, {Jansen}, {Jordi}, {Klioner}, {Lammers}, {Lindegren}, {Luri}, {Mignard}, {Panem}, {Pourbaix}, {Randich}, {Sartoretti}, {Soubiran}, {Walton}, {Arenou}, {Bailer-Jones}, {Bastian}, {Cropper}, {Drimmel}, {Katz}, {Lattanzi}, {van Leeuwen}, {Bakker}, {Cacciari}, {Casta{\~n}eda}, {De Angeli}, {Ducourant}, {Fabricius}, {Fouesneau}, {Fr{\'e}mat}, {Guerra}, {Guerrier}, {Guiraud}, {Jean-Antoine Piccolo}, {Masana}, {Messineo}, {Mowlavi}, {Nicolas}, {Nienartowicz}, {Pailler}, {Panuzzo}, {Riclet}, {Roux}, {Seabroke}, {Sordo}, {Tanga}, {Th{\'e}venin}, {Gracia-Abril}, {Portell}, {Teyssier}, {Altmann}, {Andrae}, {Bellas-Velidis}, {Benson}, {Berthier}, {Blomme}, {Brugaletta}, {Burgess}, {Busso}, {Carry}, {Cellino}, {Cheek}, {Clementini}, {Damerdji}, {Davidson}, {Delchambre}, {Dell'Oro}, {Fern{\'a}ndez-Hern{\'a}ndez}, {Galluccio},
  {Garc{\'\i}a-Lario}, {Garcia-Reinaldos}, {Gonz{\'a}lez-N{\'u}{\~n}ez}, {Gosset}, {Haigron}, {Halbwachs}, {Hambly}, {Harrison}, {Hatzidimitriou}, {Heiter}, {Hern{\'a}ndez}, {Hestroffer}, {Hodgkin}, {Holl}, {Jan{\ss}en}, {Jevardat de Fombelle}, {Jordan}, {Krone-Martins}, {Lanzafame}, {L{\"o}ffler}, {Lorca}, {Manteiga}, {Marchal}, {Marrese}, {Moitinho}, {Mora}, {Muinonen}, {Osborne}, {Pancino}, {Pauwels}, {Petit}, {Recio-Blanco}, {Richards}, {Riello}, {Rimoldini}, {Robin}, {Roegiers}, {Rybizki}, {Sarro}, {Siopis}, {Smith}, {Sozzetti}, {Ulla}, {Utrilla}, {van Leeuwen}, {van Reeven}, {Abbas}, {Abreu Aramburu}, {Accart}, {Aerts}, {Aguado}, {Ajaj}, {Altavilla}, {{\'A}lvarez}, {{\'A}lvarez Cid-Fuentes}, {Alves}, {Anderson}, {Anglada Varela}, {Antoja}, {Audard}, {Baines}, {Baker}, {Balaguer-N{\'u}{\~n}ez}, {Balbinot}, {Balog}, {Barache}, {Barbato}, {Barros}, {Barstow}, {Bartolom{\'e}}, {Bassilana}, {Bauchet}, {Baudesson-Stella}, {Becciani}, {Bellazzini}, {Bernet}, {Bertone}, {Bianchi}, {Blanco-Cuaresma}, {Boch},
  {Bombrun}, {Bossini}, {Bouquillon}, {Bragaglia}, {Bramante}, {Breedt}, {Bressan}, {Brouillet}, {Bucciarelli}, {Burlacu}, {Busonero}, {Butkevich}, {Buzzi}, {Caffau}, {Cancelliere}, {C{\'a}novas}, {Cantat-Gaudin}, {Carballo}, {Carlucci}, {Carnerero}, {Carrasco}, {Casamiquela}, {Castellani}, {Castro-Ginard}, {Castro Sampol}, {Chaoul}, {Charlot}, {Chemin}, {Chiavassa}, {Cioni}, {Comoretto}, {Cooper}, {Cornez}, {Cowell}, {Crifo}, {Crosta}, {Crowley}, {Dafonte}, {Dapergolas}, {David}, {David}, {de Laverny}, {De Luise}, {De March}, {De Ridder}, {de Souza}, {de Teodoro}, {de Torres}, {del Peloso}, {del Pozo}, {Delbo}, {Delgado}, {Delgado}, {Delisle}, {Di Matteo}, {Diakite}, {Diener}, {Distefano}, {Dolding}, {Eappachen}, {Edvardsson}, {Enke}, {Esquej}, {Fabre}, {Fabrizio}, {Faigler}, {Fedorets}, {Fernique}, {Fienga}, {Figueras}, {Fouron}, {Fragkoudi}, {Fraile}, {Franke}, {Gai}, {Garabato}, {Garcia-Gutierrez}, {Garc{\'\i}a-Torres}, {Garofalo}, {Gavras}, {Gerlach}, {Geyer}, {Giacobbe}, {Gilmore}, {Girona},
  {Giuffrida}, {Gomel}, {Gomez}, {Gonzalez-Santamaria}, {Gonz{\'a}lez-Vidal}, {Granvik}, {Guti{\'e}rrez-S{\'a}nchez}, {Guy}, {Hauser}, {Haywood}, {Helmi}, {Hidalgo}, {Hilger}, {H{\l}adczuk}, {Hobbs}, {Holland}, {Huckle}, {Jasniewicz}, {Jonker}, {Juaristi Campillo}, {Julbe}, {Karbevska}, {Kervella}, {Khanna}, {Kochoska}, {Kontizas}, {Kordopatis}, {Korn}, {Kostrzewa-Rutkowska}, {Kruszy{\'n}ska}, {Lambert}, {Lanza}, {Lasne}, {Le Campion}, {Le Fustec}, {Lebreton}, {Lebzelter}, {Leccia}, {Leclerc}, {Lecoeur-Taibi}, {Liao}, {Licata}, {Lindstr{\o}m}, {Lister}, {Livanou}, {Lobel}, {Madrero Pardo}, {Managau}, {Mann}, {Marchant}, {Marconi}, {Marcos Santos}, {Marinoni}, {Marocco}, {Marshall}, {Martin Polo}, {Mart{\'\i}n-Fleitas}, {Masip}, {Massari}, {Mastrobuono-Battisti}, {Mazeh}, {McMillan}, {Messina}, {Michalik}, {Millar}, {Mints}, {Molina}, {Molinaro}, {Moln{\'a}r}, {Montegriffo}, {Mor}, {Morbidelli}, {Morel}, {Morris}, {Mulone}, {Munoz}, {Muraveva}, {Murphy}, {Musella}, {Noval}, {Ord{\'e}novic}, {Orr{\`u}},
  {Osinde}, {Pagani}, {Pagano}, {Palaversa}, {Palicio}, {Panahi}, {Pawlak}, {Pe{\~n}alosa Esteller}, {Penttil{\"a}}, {Piersimoni}, {Pineau}, {Plachy}, {Plum}, {Poggio}, {Poretti}, {Poujoulet}, {Pr{\v{s}}a}, {Pulone}, {Racero}, {Ragaini}, {Rainer}, {Raiteri}, {Rambaux}, {Ramos}, {Ramos-Lerate}, {Re Fiorentin}, {Regibo}, {Reyl{\'e}}, {Ripepi}, {Riva}, {Rixon}, {Robichon}, {Robin}, {Roelens}, {Rohrbasser}, {Romero-G{\'o}mez}, {Rowell}, {Royer}, {Rybicki}, {Sadowski}, {Sagrist{\`a} Sell{\'e}s}, {Sahlmann}, {Salgado}, {Salguero}, {Samaras}, {Sanchez Gimenez}, {Sanna}, {Santove{\~n}a}, {Sarasso}, {Schultheis}, {Sciacca}, {Segol}, {Segovia}, {S{\'e}gransan}, {Semeux}, {Shahaf}, {Siddiqui}, {Siebert}, {Siltala}, {Slezak}, {Smart}, {Solano}, {Solitro}, {Souami}, {Souchay}, {Spagna}, {Spoto}, {Steele}, {Steidelm{\"u}ller}, {Stephenson}, {S{\"u}veges}, {Szabados}, {Szegedi-Elek}, {Taris}, {Tauran}, {Taylor}, {Teixeira}, {Thuillot}, {Tonello}, {Torra}, {Torra}, {Turon}, {Unger}, {Vaillant}, {van Dillen}, {Vanel},
  {Vecchiato}, {Viala}, {Vicente}, {Voutsinas}, {Weiler}, {Wevers}, {Wyrzykowski}, {Yoldas}, {Yvard}, {Zhao}, {Zorec}, {Zucker}, {Zurbach}, \& {Zwitter}}]{Gaia_collab2021}
{Gaia Collaboration}, {Brown}, A.~G.~A., {Vallenari}, A., {et~al.} 2021{\natexlab{a}}, \aap, 649, A1

\bibitem[{{Gaia Collaboration} {et~al.}(2021{\natexlab{b}}){Gaia Collaboration}, {Smart}, {Sarro}, {Rybizki}, {Reyl{\'e}}, {Robin}, {Hambly}, {Abbas}, {Barstow}, {de Bruijne}, {Bucciarelli}, {Carrasco}, {Cooper}, {Hodgkin}, {Masana}, {Michalik}, {Sahlmann}, {Sozzetti}, {Brown}, {Vallenari}, {Prusti}, {Babusiaux}, {Biermann}, {Creevey}, {Evans}, {Eyer}, {Hutton}, {Jansen}, {Jordi}, {Klioner}, {Lammers}, {Lindegren}, {Luri}, {Mignard}, {Panem}, {Pourbaix}, {Randich}, {Sartoretti}, {Soubiran}, {Walton}, {Arenou}, {Bailer-Jones}, {Bastian}, {Cropper}, {Drimmel}, {Katz}, {Lattanzi}, {van Leeuwen}, {Bakker}, {Casta{\~n}eda}, {De Angeli}, {Ducourant}, {Fabricius}, {Fouesneau}, {Fr{\'e}mat}, {Guerra}, {Guerrier}, {Guiraud}, {Jean-Antoine Piccolo}, {Messineo}, {Mowlavi}, {Nicolas}, {Nienartowicz}, {Pailler}, {Panuzzo}, {Riclet}, {Roux}, {Seabroke}, {Sordo}, {Tanga}, {Th{\'e}venin}, {Gracia-Abril}, {Portell}, {Teyssier}, {Altmann}, {Andrae}, {Bellas-Velidis}, {Benson}, {Berthier}, {Blomme}, {Brugaletta}, {Burgess},
  {Busso}, {Carry}, {Cellino}, {Cheek}, {Clementini}, {Damerdji}, {Davidson}, {Delchambre}, {Dell'Oro}, {Fern{\'a}ndez-Hern{\'a}ndez}, {Galluccio}, {Garc{\'\i}a-Lario}, {Garcia-Reinaldos}, {Gonz{\'a}lez-N{\'u}{\~n}ez}, {Gosset}, {Haigron}, {Halbwachs}, {Harrison}, {Hatzidimitriou}, {Heiter}, {Hern{\'a}ndez}, {Hestroffer}, {Holl}, {Jan{\ss}en}, {Jevardat de Fombelle}, {Jordan}, {Krone-Martins}, {Lanzafame}, {L{\"o}ffler}, {Lorca}, {Manteiga}, {Marchal}, {Marrese}, {Moitinho}, {Mora}, {Muinonen}, {Osborne}, {Pancino}, {Pauwels}, {Recio-Blanco}, {Richards}, {Riello}, {Rimoldini}, {Roegiers}, {Siopis}, {Smith}, {Ulla}, {Utrilla}, {van Leeuwen}, {van Reeven}, {Abreu Aramburu}, {Accart}, {Aerts}, {Aguado}, {Ajaj}, {Altavilla}, {{\'A}lvarez}, {{\'A}lvarez Cid-Fuentes}, {Alves}, {Anderson}, {Anglada Varela}, {Antoja}, {Audard}, {Baines}, {Baker}, {Balaguer-N{\'u}{\~n}ez}, {Balbinot}, {Balog}, {Barache}, {Barbato}, {Barros}, {Bartolom{\'e}}, {Bassilana}, {Bauchet}, {Baudesson-Stella}, {Becciani}, {Bellazzini},
  {Bernet}, {Bertone}, {Bianchi}, {Blanco-Cuaresma}, {Boch}, {Bombrun}, {Bossini}, {Bouquillon}, {Bragaglia}, {Bramante}, {Breedt}, {Bressan}, {Brouillet}, {Burlacu}, {Busonero}, {Butkevich}, {Buzzi}, {Caffau}, {Cancelliere}, {C{\'a}novas}, {Cantat-Gaudin}, {Carballo}, {Carlucci}, {Carnerero}, {Casamiquela}, {Castellani}, {Castro-Ginard}, {Castro Sampol}, {Chaoul}, {Charlot}, {Chemin}, {Chiavassa}, {Cioni}, {Comoretto}, {Cornez}, {Cowell}, {Crifo}, {Crosta}, {Crowley}, {Dafonte}, {Dapergolas}, {David}, {David}, {de Laverny}, {De Luise}, {De March}, {De Ridder}, {de Souza}, {de Teodoro}, {de Torres}, {del Peloso}, {del Pozo}, {Delgado}, {Delgado}, {Delisle}, {Di Matteo}, {Diakite}, {Diener}, {Distefano}, {Dolding}, {Eappachen}, {Edvardsson}, {Enke}, {Esquej}, {Fabre}, {Fabrizio}, {Faigler}, {Fedorets}, {Fernique}, {Fienga}, {Figueras}, {Fouron}, {Fragkoudi}, {Fraile}, {Franke}, {Gai}, {Garabato}, {Garcia-Gutierrez}, {Garc{\'\i}a-Torres}, {Garofalo}, {Gavras}, {Gerlach}, {Geyer}, {Giacobbe}, {Gilmore},
  {Girona}, {Giuffrida}, {Gomel}, {Gomez}, {Gonzalez-Santamaria}, {Gonz{\'a}lez-Vidal}, {Granvik}, {Guti{\'e}rrez-S{\'a}nchez}, {Guy}, {Hauser}, {Haywood}, {Helmi}, {Hidalgo}, {Hilger}, {H{\l}adczuk}, {Hobbs}, {Holland}, {Huckle}, {Jasniewicz}, {Jonker}, {Juaristi Campillo}, {Julbe}, {Karbevska}, {Kervella}, {Khanna}, {Kochoska}, {Kontizas}, {Kordopatis}, {Korn}, {Kostrzewa-Rutkowska}, {Kruszy{\'n}ska}, {Lambert}, {Lanza}, {Lasne}, {Le Campion}, {Le Fustec}, {Lebreton}, {Lebzelter}, {Leccia}, {Leclerc}, {Lecoeur-Taibi}, {Liao}, {Licata}, {Lindstr{\o}m}, {Lister}, {Livanou}, {Lobel}, {Madrero Pardo}, {Managau}, {Mann}, {Marchant}, {Marconi}, {Marcos Santos}, {Marinoni}, {Marocco}, {Marshall}, {Martin Polo}, {Mart{\'\i}n-Fleitas}, {Masip}, {Massari}, {Mastrobuono-Battisti}, {Mazeh}, {McMillan}, {Messina}, {Millar}, {Mints}, {Molina}, {Molinaro}, {Moln{\'a}r}, {Montegriffo}, {Mor}, {Morbidelli}, {Morel}, {Morris}, {Mulone}, {Munoz}, {Muraveva}, {Murphy}, {Musella}, {Noval}, {Ord{\'e}novic}, {Orr{\`u}}, {Osinde},
  {Pagani}, {Pagano}, {Palaversa}, {Palicio}, {Panahi}, {Pawlak}, {Pe{\~n}alosa Esteller}, {Penttil{\"a}}, {Piersimoni}, {Pineau}, {Plachy}, {Plum}, {Poggio}, {Poretti}, {Poujoulet}, {Pr{\v{s}}a}, {Pulone}, {Racero}, {Ragaini}, {Rainer}, {Raiteri}, {Rambaux}, {Ramos}, {Ramos-Lerate}, {Re Fiorentin}, {Regibo}, {Ripepi}, {Riva}, {Rixon}, {Robichon}, {Robin}, {Roelens}, {Rohrbasser}, {Romero-G{\'o}mez}, {Rowell}, {Royer}, {Rybicki}, {Sadowski}, {Sagrist{\`a} Sell{\'e}s}, {Salgado}, {Salguero}, {Samaras}, {Sanchez Gimenez}, {Sanna}, {Santove{\~n}a}, {Sarasso}, {Schultheis}, {Sciacca}, {Segol}, {Segovia}, {S{\'e}gransan}, {Semeux}, {Shahaf}, {Siddiqui}, {Siebert}, {Siltala}, {Slezak}, {Solano}, {Solitro}, {Souami}, {Souchay}, {Spagna}, {Spoto}, {Steele}, {Steidelm{\"u}ller}, {Stephenson}, {S{\"u}veges}, {Szabados}, {Szegedi-Elek}, {Taris}, {Tauran}, {Taylor}, {Teixeira}, {Thuillot}, {Tonello}, {Torra}, {Torra}, {Turon}, {Unger}, {Vaillant}, {van Dillen}, {Vanel}, {Vecchiato}, {Viala}, {Vicente}, {Voutsinas},
  {Weiler}, {Wevers}, {Wyrzykowski}, {Yoldas}, {Yvard}, {Zhao}, {Zorec}, {Zucker}, {Zurbach}, \& {Zwitter}}]{GaiaCollaboration2021}
{Gaia Collaboration}, {Smart}, R.~L., {Sarro}, L.~M., {et~al.} 2021{\natexlab{b}}, \aap, 649, A6

\bibitem[{{Gaia Collaboration} {et~al.}(2023){Gaia Collaboration}, {Vallenari}, {Brown}, {Prusti}, {de Bruijne}, {Arenou}, {Babusiaux}, {Biermann}, {Creevey}, {Ducourant}, {Evans}, {Eyer}, {Guerra}, {Hutton}, {Jordi}, {Klioner}, {Lammers}, {Lindegren}, {Luri}, {Mignard}, {Panem}, {Pourbaix}, {Randich}, {Sartoretti}, {Soubiran}, {Tanga}, {Walton}, {Bailer-Jones}, {Bastian}, {Drimmel}, {Jansen}, {Katz}, {Lattanzi}, {van Leeuwen}, {Bakker}, {Cacciari}, {Casta{\~n}eda}, {De Angeli}, {Fabricius}, {Fouesneau}, {Fr{\'e}mat}, {Galluccio}, {Guerrier}, {Heiter}, {Masana}, {Messineo}, {Mowlavi}, {Nicolas}, {Nienartowicz}, {Pailler}, {Panuzzo}, {Riclet}, {Roux}, {Seabroke}, {Sordo}, {Th{\'e}venin}, {Gracia-Abril}, {Portell}, {Teyssier}, {Altmann}, {Andrae}, {Audard}, {Bellas-Velidis}, {Benson}, {Berthier}, {Blomme}, {Burgess}, {Busonero}, {Busso}, {C{\'a}novas}, {Carry}, {Cellino}, {Cheek}, {Clementini}, {Damerdji}, {Davidson}, {de Teodoro}, {Nu{\~n}ez Campos}, {Delchambre}, {Dell'Oro}, {Esquej},
  {Fern{\'a}ndez-Hern{\'a}ndez}, {Fraile}, {Garabato}, {Garc{\'\i}a-Lario}, {Gosset}, {Haigron}, {Halbwachs}, {Hambly}, {Harrison}, {Hern{\'a}ndez}, {Hestroffer}, {Hodgkin}, {Holl}, {Jan{\ss}en}, {Jevardat de Fombelle}, {Jordan}, {Krone-Martins}, {Lanzafame}, {L{\"o}ffler}, {Marchal}, {Marrese}, {Moitinho}, {Muinonen}, {Osborne}, {Pancino}, {Pauwels}, {Recio-Blanco}, {Reyl{\'e}}, {Riello}, {Rimoldini}, {Roegiers}, {Rybizki}, {Sarro}, {Siopis}, {Smith}, {Sozzetti}, {Utrilla}, {van Leeuwen}, {Abbas}, {{\'A}brah{\'a}m}, {Abreu Aramburu}, {Aerts}, {Aguado}, {Ajaj}, {Aldea-Montero}, {Altavilla}, {{\'A}lvarez}, {Alves}, {Anders}, {Anderson}, {Anglada Varela}, {Antoja}, {Baines}, {Baker}, {Balaguer-N{\'u}{\~n}ez}, {Balbinot}, {Balog}, {Barache}, {Barbato}, {Barros}, {Barstow}, {Bartolom{\'e}}, {Bassilana}, {Bauchet}, {Becciani}, {Bellazzini}, {Berihuete}, {Bernet}, {Bertone}, {Bianchi}, {Binnenfeld}, {Blanco-Cuaresma}, {Blazere}, {Boch}, {Bombrun}, {Bossini}, {Bouquillon}, {Bragaglia}, {Bramante}, {Breedt},
  {Bressan}, {Brouillet}, {Brugaletta}, {Bucciarelli}, {Burlacu}, {Butkevich}, {Buzzi}, {Caffau}, {Cancelliere}, {Cantat-Gaudin}, {Carballo}, {Carlucci}, {Carnerero}, {Carrasco}, {Casamiquela}, {Castellani}, {Castro-Ginard}, {Chaoul}, {Charlot}, {Chemin}, {Chiaramida}, {Chiavassa}, {Chornay}, {Comoretto}, {Contursi}, {Cooper}, {Cornez}, {Cowell}, {Crifo}, {Cropper}, {Crosta}, {Crowley}, {Dafonte}, {Dapergolas}, {David}, {David}, {de Laverny}, {De Luise}, {De March}, {De Ridder}, {de Souza}, {de Torres}, {del Peloso}, {del Pozo}, {Delbo}, {Delgado}, {Delisle}, {Demouchy}, {Dharmawardena}, {Di Matteo}, {Diakite}, {Diener}, {Distefano}, {Dolding}, {Edvardsson}, {Enke}, {Fabre}, {Fabrizio}, {Faigler}, {Fedorets}, {Fernique}, {Fienga}, {Figueras}, {Fournier}, {Fouron}, {Fragkoudi}, {Gai}, {Garcia-Gutierrez}, {Garcia-Reinaldos}, {Garc{\'\i}a-Torres}, {Garofalo}, {Gavel}, {Gavras}, {Gerlach}, {Geyer}, {Giacobbe}, {Gilmore}, {Girona}, {Giuffrida}, {Gomel}, {Gomez}, {Gonz{\'a}lez-N{\'u}{\~n}ez},
  {Gonz{\'a}lez-Santamar{\'\i}a}, {Gonz{\'a}lez-Vidal}, {Granvik}, {Guillout}, {Guiraud}, {Guti{\'e}rrez-S{\'a}nchez}, {Guy}, {Hatzidimitriou}, {Hauser}, {Haywood}, {Helmer}, {Helmi}, {Sarmiento}, {Hidalgo}, {Hilger}, {H{\l}adczuk}, {Hobbs}, {Holland}, {Huckle}, {Jardine}, {Jasniewicz}, {Jean-Antoine Piccolo}, {Jim{\'e}nez-Arranz}, {Jorissen}, {Juaristi Campillo}, {Julbe}, {Karbevska}, {Kervella}, {Khanna}, {Kontizas}, {Kordopatis}, {Korn}, {K{\'o}sp{\'a}l}, {Kostrzewa-Rutkowska}, {Kruszy{\'n}ska}, {Kun}, {Laizeau}, {Lambert}, {Lanza}, {Lasne}, {Le Campion}, {Lebreton}, {Lebzelter}, {Leccia}, {Leclerc}, {Lecoeur-Taibi}, {Liao}, {Licata}, {Lindstr{\o}m}, {Lister}, {Livanou}, {Lobel}, {Lorca}, {Loup}, {Madrero Pardo}, {Magdaleno Romeo}, {Managau}, {Mann}, {Manteiga}, {Marchant}, {Marconi}, {Marcos}, {Marcos Santos}, {Mar{\'\i}n Pina}, {Marinoni}, {Marocco}, {Marshall}, {Martin Polo}, {Mart{\'\i}n-Fleitas}, {Marton}, {Mary}, {Masip}, {Massari}, {Mastrobuono-Battisti}, {Mazeh}, {McMillan}, {Messina}, {Michalik},
  {Millar}, {Mints}, {Molina}, {Molinaro}, {Moln{\'a}r}, {Monari}, {Mongui{\'o}}, {Montegriffo}, {Montero}, {Mor}, {Mora}, {Morbidelli}, {Morel}, {Morris}, {Muraveva}, {Murphy}, {Musella}, {Nagy}, {Noval}, {Oca{\~n}a}, {Ogden}, {Ordenovic}, {Osinde}, {Pagani}, {Pagano}, {Palaversa}, {Palicio}, {Pallas-Quintela}, {Panahi}, {Payne-Wardenaar}, {Pe{\~n}alosa Esteller}, {Penttil{\"a}}, {Pichon}, {Piersimoni}, {Pineau}, {Plachy}, {Plum}, {Poggio}, {Pr{\v{s}}a}, {Pulone}, {Racero}, {Ragaini}, {Rainer}, {Raiteri}, {Rambaux}, {Ramos}, {Ramos-Lerate}, {Re Fiorentin}, {Regibo}, {Richards}, {Rios Diaz}, {Ripepi}, {Riva}, {Rix}, {Rixon}, {Robichon}, {Robin}, {Robin}, {Roelens}, {Rogues}, {Rohrbasser}, {Romero-G{\'o}mez}, {Rowell}, {Royer}, {Ruz Mieres}, {Rybicki}, {Sadowski}, {S{\'a}ez N{\'u}{\~n}ez}, {Sagrist{\`a} Sell{\'e}s}, {Sahlmann}, {Salguero}, {Samaras}, {Sanchez Gimenez}, {Sanna}, {Santove{\~n}a}, {Sarasso}, {Schultheis}, {Sciacca}, {Segol}, {Segovia}, {S{\'e}gransan}, {Semeux}, {Shahaf}, {Siddiqui}, {Siebert},
  {Siltala}, {Silvelo}, {Slezak}, {Slezak}, {Smart}, {Snaith}, {Solano}, {Solitro}, {Souami}, {Souchay}, {Spagna}, {Spina}, {Spoto}, {Steele}, {Steidelm{\"u}ller}, {Stephenson}, {S{\"u}veges}, {Surdej}, {Szabados}, {Szegedi-Elek}, {Taris}, {Taylor}, {Teixeira}, {Tolomei}, {Tonello}, {Torra}, {Torra}, {Torralba Elipe}, {Trabucchi}, {Tsounis}, {Turon}, {Ulla}, {Unger}, {Vaillant}, {van Dillen}, {van Reeven}, {Vanel}, {Vecchiato}, {Viala}, {Vicente}, {Voutsinas}, {Weiler}, {Wevers}, {Wyrzykowski}, {Yoldas}, {Yvard}, {Zhao}, {Zorec}, {Zucker}, \& {Zwitter}}]{GaiaCollab2023}
{Gaia Collaboration}, {Vallenari}, A., {Brown}, A.~G.~A., {et~al.} 2023, \aap, 674, A1

\bibitem[{{Geier}(2020)}]{Geier2020}
{Geier}, S. 2020, \aap, 635, A193

\bibitem[{{Geier} {et~al.}(2023){Geier}, {Dorsch}, {Dawson}, {Pelisoli}, {Munday}, {Marsh}, {Schaffenroth}, \& {Heber}}]{Geier2023}
{Geier}, S., {Dorsch}, M., {Dawson}, H., {et~al.} 2023, \aap, 677, A11

\bibitem[{{Geier} {et~al.}(2022){Geier}, {Dorsch}, {Pelisoli}, {Reindl}, {Heber}, \& {Irrgang}}]{Geier2022}
{Geier}, S., {Dorsch}, M., {Pelisoli}, I., {et~al.} 2022, \aap, 661, A113

\bibitem[{{Geier} {et~al.}(2010){Geier}, {Heber}, {Tillich}, {Hirsch}, {Edelmann}, {Schaffenroth}, {Kupfer}, {M{\"u}ller}, {Maxted}, {{\O}stensen}, {Podsiadlowski}, {Marsh}, {G{\"a}nsicke}, {Morales-Rueda}, {Nelemans}, {Napiwotzki}, {G{\"u}nther}, \& {Carone}}]{Geier2010}
{Geier}, S., {Heber}, U., {Tillich}, A., {et~al.} 2010, \apss, 329, 91

\bibitem[{{Geier} {et~al.}(2019){Geier}, {Raddi}, {Gentile Fusillo}, \& {Marsh}}]{Geier2019}
{Geier}, S., {Raddi}, R., {Gentile Fusillo}, N.~P., \& {Marsh}, T.~R. 2019, \aap, 621, A38

\bibitem[{{Ginsburg} {et~al.}(2019){Ginsburg}, {Sip{\H{o}}cz}, {Brasseur}, {Cowperthwaite}, {Craig}, {Deil}, {Guillochon}, {Guzman}, {Liedtke}, {Lian Lim}, {Lockhart}, {Mommert}, {Morris}, {Norman}, {Parikh}, {Persson}, {Robitaille}, {Segovia}, {Singer}, {Tollerud}, {de Val-Borro}, {Valtchanov}, {Woillez}, {Astroquery Collaboration}, \& {a subset of astropy Collaboration}}]{Ginsburg2019}
{Ginsburg}, A., {Sip{\H{o}}cz}, B.~M., {Brasseur}, C.~E., {et~al.} 2019, \aj, 157, 98

\bibitem[{{Groot} {et~al.}(2024){Groot}, {Bloemen}, {Vreeswijk}, {van Roestel}, {Jonker}, {Nelemans}, {Klein-Wolt}, {Lepoole}, {Pieterse}, {Rodenhuis}, {Boland}, {Haverkorn}, {Aerts}, {Bakker}, {Balster}, {Bekema}, {Dijkstra}, {Dolron}, {Elswijk}, {van Elteren}, {Engels}, {Fokker}, {de Haan}, {Hahn}, {ter Horst}, {Lesman}, {Kragt}, {Morren}, {Nillissen}, {Pessemier}, {Raskin}, {de Rijke}, {Scheers}, {Schuil}, {Timmer}, {Antunes Amaral}, {Arancibia-Rojas}, {Arcavi}, {Blagorodnova}, {Biswas}, {Breton}, {Dawson}, {Dayal}, {De Wet}, {Duffy}, {Faris}, {Fausnaugh}, {Gal-Yam}, {Geier}, {Horesh}, {Johnston}, {Katusiime}, {Kelley}, {Kosakowski}, {Kupfer}, {Leloudas}, {Levan}, {Modiano}, {Mogawana}, {Munday}, {Paice}, {Patat}, {Pelisoli}, {Ramsay}, {Ranaivomanana}, {Ruiz-Carmona}, {Schaffenroth}, {Scaringi}, {Stoppa}, {Street}, {Tranin}, {Uzundag}, {Valenti}, {Veresvarska}, {Vuc̆kovi{\'c}}, {Wichern}, {Wijers}, {Wijnands}, \& {Zimmerman}}]{Groot2024}
{Groot}, P.~J., {Bloemen}, S., {Vreeswijk}, P.~M., {et~al.} 2024, \pasp, 136, 115003

\bibitem[{{Han} {et~al.}(2003){Han}, {Podsiadlowski}, {Maxted}, \& {Marsh}}]{Han2003}
{Han}, Z., {Podsiadlowski}, P., {Maxted}, P.~F.~L., \& {Marsh}, T.~R. 2003, \mnras, 341, 669

\bibitem[{{Han} {et~al.}(2002){Han}, {Podsiadlowski}, {Maxted}, {Marsh}, \& {Ivanova}}]{Han2002}
{Han}, Z., {Podsiadlowski}, P., {Maxted}, P.~F.~L., {Marsh}, T.~R., \& {Ivanova}, N. 2002, \mnras, 336, 449

\bibitem[{{Heber}(2009)}]{Heber2009}
{Heber}, U. 2009, \araa, 47, 211

\bibitem[{{Heber}(2016)}]{Heber2016}
{Heber}, U. 2016, \pasp, 128, 082001

\bibitem[{{Hou} {et~al.}(2023){Hou}, {Luo}, {Dong}, {Chen}, \& {Bai}}]{Hou2023}
{Hou}, W., {Luo}, A.~L., {Dong}, Y.-Q., {Chen}, X.-L., \& {Bai}, Z.-R. 2023, \aj, 165, 148

\bibitem[{{Hu} {et~al.}(2008){Hu}, {Dupret}, {Aerts}, {Nelemans}, {Kawaler}, {Miglio}, {Montalban}, \& {Scuflaire}}]{Hu2008}
{Hu}, H., {Dupret}, M.~A., {Aerts}, C., {et~al.} 2008, \aap, 490, 243

\bibitem[{{Hu} {et~al.}(2011){Hu}, {Tout}, {Glebbeek}, \& {Dupret}}]{Hu2011}
{Hu}, H., {Tout}, C.~A., {Glebbeek}, E., \& {Dupret}, M.~A. 2011, \mnras, 418, 195

\bibitem[{{Ivezi{\'c}} {et~al.}(2019){Ivezi{\'c}}, {Kahn}, {Tyson}, {Abel}, {Acosta}, {Allsman}, {Alonso}, {AlSayyad}, {Anderson}, {Andrew}, {Angel}, {Angeli}, {Ansari}, {Antilogus}, {Araujo}, {Armstrong}, {Arndt}, {Astier}, {Aubourg}, {Auza}, {Axelrod}, {Bard}, {Barr}, {Barrau}, {Bartlett}, {Bauer}, {Bauman}, {Baumont}, {Bechtol}, {Bechtol}, {Becker}, {Becla}, {Beldica}, {Bellavia}, {Bianco}, {Biswas}, {Blanc}, {Blazek}, {Blandford}, {Bloom}, {Bogart}, {Bond}, {Booth}, {Borgland}, {Borne}, {Bosch}, {Boutigny}, {Brackett}, {Bradshaw}, {Brandt}, {Brown}, {Bullock}, {Burchat}, {Burke}, {Cagnoli}, {Calabrese}, {Callahan}, {Callen}, {Carlin}, {Carlson}, {Chandrasekharan}, {Charles-Emerson}, {Chesley}, {Cheu}, {Chiang}, {Chiang}, {Chirino}, {Chow}, {Ciardi}, {Claver}, {Cohen-Tanugi}, {Cockrum}, {Coles}, {Connolly}, {Cook}, {Cooray}, {Covey}, {Cribbs}, {Cui}, {Cutri}, {Daly}, {Daniel}, {Daruich}, {Daubard}, {Daues}, {Dawson}, {Delgado}, {Dellapenna}, {de Peyster}, {de Val-Borro}, {Digel}, {Doherty}, {Dubois},
  {Dubois-Felsmann}, {Durech}, {Economou}, {Eifler}, {Eracleous}, {Emmons}, {Fausti Neto}, {Ferguson}, {Figueroa}, {Fisher-Levine}, {Focke}, {Foss}, {Frank}, {Freemon}, {Gangler}, {Gawiser}, {Geary}, {Gee}, {Geha}, {Gessner}, {Gibson}, {Gilmore}, {Glanzman}, {Glick}, {Goldina}, {Goldstein}, {Goodenow}, {Graham}, {Gressler}, {Gris}, {Guy}, {Guyonnet}, {Haller}, {Harris}, {Hascall}, {Haupt}, {Hernandez}, {Herrmann}, {Hileman}, {Hoblitt}, {Hodgson}, {Hogan}, {Howard}, {Huang}, {Huffer}, {Ingraham}, {Innes}, {Jacoby}, {Jain}, {Jammes}, {Jee}, {Jenness}, {Jernigan}, {Jevremovi{\'c}}, {Johns}, {Johnson}, {Johnson}, {Jones}, {Juramy-Gilles}, {Juri{\'c}}, {Kalirai}, {Kallivayalil}, {Kalmbach}, {Kantor}, {Karst}, {Kasliwal}, {Kelly}, {Kessler}, {Kinnison}, {Kirkby}, {Knox}, {Kotov}, {Krabbendam}, {Krughoff}, {Kub{\'a}nek}, {Kuczewski}, {Kulkarni}, {Ku}, {Kurita}, {Lage}, {Lambert}, {Lange}, {Langton}, {Le Guillou}, {Levine}, {Liang}, {Lim}, {Lintott}, {Long}, {Lopez}, {Lotz}, {Lupton}, {Lust}, {MacArthur}, {Mahabal},
  {Mandelbaum}, {Markiewicz}, {Marsh}, {Marshall}, {Marshall}, {May}, {McKercher}, {McQueen}, {Meyers}, {Migliore}, {Miller}, {Mills}, {Miraval}, {Moeyens}, {Moolekamp}, {Monet}, {Moniez}, {Monkewitz}, {Montgomery}, {Morrison}, {Mueller}, {Muller}, {Mu{\~n}oz Arancibia}, {Neill}, {Newbry}, {Nief}, {Nomerotski}, {Nordby}, {O'Connor}, {Oliver}, {Olivier}, {Olsen}, {O'Mullane}, {Ortiz}, {Osier}, {Owen}, {Pain}, {Palecek}, {Parejko}, {Parsons}, {Pease}, {Peterson}, {Peterson}, {Petravick}, {Libby Petrick}, {Petry}, {Pierfederici}, {Pietrowicz}, {Pike}, {Pinto}, {Plante}, {Plate}, {Plutchak}, {Price}, {Prouza}, {Radeka}, {Rajagopal}, {Rasmussen}, {Regnault}, {Reil}, {Reiss}, {Reuter}, {Ridgway}, {Riot}, {Ritz}, {Robinson}, {Roby}, {Roodman}, {Rosing}, {Roucelle}, {Rumore}, {Russo}, {Saha}, {Sassolas}, {Schalk}, {Schellart}, {Schindler}, {Schmidt}, {Schneider}, {Schneider}, {Schoening}, {Schumacher}, {Schwamb}, {Sebag}, {Selvy}, {Sembroski}, {Seppala}, {Serio}, {Serrano}, {Shaw}, {Shipsey}, {Sick}, {Silvestri},
  {Slater}, {Smith}, {Smith}, {Sobhani}, {Soldahl}, {Storrie-Lombardi}, {Stover}, {Strauss}, {Street}, {Stubbs}, {Sullivan}, {Sweeney}, {Swinbank}, {Szalay}, {Takacs}, {Tether}, {Thaler}, {Thayer}, {Thomas}, {Thornton}, {Thukral}, {Tice}, {Trilling}, {Turri}, {Van Berg}, {Vanden Berk}, {Vetter}, {Virieux}, {Vucina}, {Wahl}, {Walkowicz}, {Walsh}, {Walter}, {Wang}, {Wang}, {Warner}, {Wiecha}, {Willman}, {Winters}, {Wittman}, {Wolff}, {Wood-Vasey}, {Wu}, {Xin}, {Yoachim}, \& {Zhan}}]{Ivezic2019}
{Ivezi{\'c}}, {\v{Z}}., {Kahn}, S.~M., {Tyson}, J.~A., {et~al.} 2019, \apj, 873, 111

\bibitem[{{Jin} {et~al.}(2024){Jin}, {Trager}, {Dalton}, {Aguerri}, {Drew}, {Falc{\'o}n-Barroso}, {G{\"a}nsicke}, {Hill}, {Iovino}, {Pieri}, {Poggianti}, {Smith}, {Vallenari}, {Abrams}, {Aguado}, {Antoja}, {Arag{\'o}n-Salamanca}, {Ascasibar}, {Babusiaux}, {Balcells}, {Barrena}, {Battaglia}, {Belokurov}, {Bensby}, {Bonifacio}, {Bragaglia}, {Carrasco}, {Carrera}, {Cornwell}, {Dom{\'\i}nguez-Palmero}, {Duncan}, {Famaey}, {Fari{\~n}a}, {Gonzalez}, {Guest}, {Hatch}, {Hess}, {Hoskin}, {Irwin}, {Knapen}, {Koposov}, {Kuchner}, {Laigle}, {Lewis}, {Longhetti}, {Lucatello}, {M{\'e}ndez-Abreu}, {Mercurio}, {Molaeinezhad}, {Mongui{\'o}}, {Morrison}, {Murphy}, {Peralta de Arriba}, {P{\'e}rez}, {P{\'e}rez-R{\`a}fols}, {Pic{\'o}}, {Raddi}, {Romero-G{\'o}mez}, {Royer}, {Siebert}, {Seabroke}, {Som}, {Terrett}, {Thomas}, {Wesson}, {Worley}, {Alfaro}, {Allende Prieto}, {Alonso-Santiago}, {Amos}, {Ashley}, {Balaguer-N{\'u}{\~n}ez}, {Balbinot}, {Bellazzini}, {Benn}, {Berlanas}, {Bernard}, {Best}, {Bettoni}, {Bianco}, {Bishop},
  {Blomqvist}, {Boeche}, {Bolzonella}, {Bonoli}, {Bosma}, {Britavskiy}, {Busarello}, {Caffau}, {Cantat-Gaudin}, {Castro-Ginard}, {Couto}, {Carbajo-Hijarrubia}, {Carter}, {Casamiquela}, {Conrado}, {Corcho-Caballero}, {Costantin}, {Deason}, {de Burgos}, {De Grandi}, {Di Matteo}, {Dom{\'\i}nguez-G{\'o}mez}, {Dorda}, {Drake}, {Dutta}, {Erkal}, {Feltzing}, {Ferr{\'e}-Mateu}, {Feuillet}, {Figueras}, {Fossati}, {Franciosini}, {Frasca}, {Fumagalli}, {Gallazzi}, {Garc{\'\i}a-Benito}, {Gentile Fusillo}, {Gebran}, {Gilbert}, {Gledhill}, {Gonz{\'a}lez Delgado}, {Greimel}, {Guarcello}, {Guerra}, {Gullieuszik}, {Haines}, {Hardcastle}, {Harris}, {Haywood}, {Helmi}, {Hernandez}, {Herrero}, {Hughes}, {Ir{\v{s}}i{\v{c}}}, {Jablonka}, {Jarvis}, {Jordi}, {Kondapally}, {Kordopatis}, {Krogager}, {La Barbera}, {Lam}, {Larsen}, {Lemasle}, {Lewis}, {Lhom{\'e}}, {Lind}, {Lodi}, {Longobardi}, {Lonoce}, {Magrini}, {Ma{\'\i}z Apell{\'a}niz}, {Marchal}, {Marco}, {Martin}, {Matsuno}, {Maurogordato}, {Merluzzi}, {Miralda-Escud{\'e}},
  {Molinari}, {Monari}, {Morelli}, {Mottram}, {Naylor}, {Negueruela}, {O{\~n}orbe}, {Pancino}, {Peirani}, {Peletier}, {Pozzetti}, {Rainer}, {Ramos}, {Read}, {Rossi}, {R{\"o}ttgering}, {Rubi{\~n}o-Mart{\'\i}n}, {Sabater}, {San Juan}, {Sanna}, {Schallig}, {Schiavon}, {Schultheis}, {Serra}, {Shimwell}, {Sim{\'o}n-D{\'\i}az}, {Smith}, {Sordo}, {Sorini}, {Soubiran}, {Starkenburg}, {Steele}, {Stott}, {Stuik}, {Tolstoy}, {Tortora}, {Tsantaki}, {Van der Swaelmen}, {van Weeren}, {Vergani}, {Verheijen}, {Verro}, {Vink}, {Vioque}, {Walcher}, {Walton}, {Wegg}, {Weijmans}, {Williams}, {Wilson}, {Wright}, {Xylakis-Dornbusch}, {Youakim}, {Zibetti}, \& {Zurita}}]{Jin2024}
{Jin}, S., {Trager}, S.~C., {Dalton}, G.~B., {et~al.} 2024, \mnras, 530, 2688

\bibitem[{{Kao} {et~al.}(2024){Kao}, {Zhang}, \& {Wu}}]{Kao2024}
{Kao}, W.-B., {Zhang}, Y., \& {Wu}, X.-B. 2024, \pasj, 76, 653

\bibitem[{{Kim} {et~al.}(2021){Kim}, {Yeo}, {Bailer-Jones}, \& {Lee}}]{Kim2021}
{Kim}, D.-W., {Yeo}, D., {Bailer-Jones}, C. A.~L., \& {Lee}, G. 2021, \aap, 653, A22

\bibitem[{{Kollmeier} {et~al.}(2019){Kollmeier}, {Anderson}, {Blanc}, {Blanton}, {Covey}, {Crane}, {Drory}, {Frinchaboy}, {Froning}, {Johnson}, {Kneib}, {Kreckel}, {Merloni}, {Pellegrini}, {Pogge}, {Ramirez}, {Rix}, {Sayres}, {S{\'a}nchez-Gallego}, {Shen}, {Tkachenko}, {Trump}, {Tuttle}, {Weijmans}, {Zasowski}, {Barbuy}, {Beaton}, {Bergemann}, {Bochanski}, {Brandt}, {Casey}, {Cherinka}, {Eracleous}, {Fan}, {Garc{\'\i}a}, {Green}, {Hekker}, {Lane}, {Longa-Pe{\~n}a}, {Mathur}, {Meza}, {Minchev}, {Myers}, {Nidever}, {Nitschelm}, {O'Connell}, {Price-Whelan}, {Raddick}, {Rossi}, {Sankrit}, {Simon}, {Stutz}, {Ting}, {Trakhtenbrot}, {Weaver}, {Willmer}, \& {Weinberg}}]{Kollmeier2019}
{Kollmeier}, J., {Anderson}, S.~F., {Blanc}, G.~A., {et~al.} 2019, in Bulletin of the American Astronomical Society, Vol.~51, 274

\bibitem[{{Krzesinski} \& {Balona}(2022)}]{Krzesinski2022}
{Krzesinski}, J. \& {Balona}, L.~A. 2022, \aap, 663, A45

\bibitem[{Kuhn \& Johnson(2019)}]{kuhn2019}
Kuhn, M. \& Johnson, K. 2019, Feature Engineering and Selection: A Practical Approach for Predictive Models (Boca Raton, FL: Chapman and Hall/CRC)

\bibitem[{{Kupfer} {et~al.}(2015){Kupfer}, {Geier}, {Heber}, {{\O}stensen}, {Barlow}, {Maxted}, {Heuser}, {Schaffenroth}, \& {G{\"a}nsicke}}]{Kupfer2015}
{Kupfer}, T., {Geier}, S., {Heber}, U., {et~al.} 2015, \aap, 576, A44

\bibitem[{{Lei} {et~al.}(2023){Lei}, {He}, {N{\'e}meth}, {Vos}, {Zou}, {Hu}, {Xiao}, {Yan}, \& {Zhao}}]{Lei2023}
{Lei}, Z., {He}, R., {N{\'e}meth}, P., {et~al.} 2023, \apj, 942, 109

\bibitem[{{Lei} {et~al.}(2020){Lei}, {Zhao}, {N{\'e}meth}, \& {Zhao}}]{Lei2020}
{Lei}, Z., {Zhao}, J., {N{\'e}meth}, P., \& {Zhao}, G. 2020, \apj, 889, 117

\bibitem[{{Liao} {et~al.}(2024){Liao}, {Ren}, {Chen}, {Li}, \& {Li}}]{Liao2024}
{Liao}, H., {Ren}, G., {Chen}, X., {Li}, Y., \& {Li}, G. 2024, \aj, 167, 180

\bibitem[{{Lightkurve Collaboration} {et~al.}(2018){Lightkurve Collaboration}, {Cardoso}, {Hedges}, {Gully-Santiago}, {Saunders}, {Cody}, {Barclay}, {Hall}, {Sagear}, {Turtelboom}, {Zhang}, {Tzanidakis}, {Mighell}, {Coughlin}, {Bell}, {Berta-Thompson}, {Williams}, {Dotson}, \& {Barentsen}}]{LC_collab2018}
{Lightkurve Collaboration}, {Cardoso}, J. V. d.~M., {Hedges}, C., {et~al.} 2018, {Lightkurve: Kepler and TESS time series analysis in Python}, Astrophysics Source Code Library, record ascl:1812.013

\bibitem[{{Luo} {et~al.}(2019){Luo}, {N{\'e}meth}, {Deng}, \& {Han}}]{Luo2019}
{Luo}, Y., {N{\'e}meth}, P., {Deng}, L., \& {Han}, Z. 2019, \apj, 881, 7

\bibitem[{{Macfarlane} {et~al.}(2015){Macfarlane}, {Toma}, {Ramsay}, {Groot}, {Woudt}, {Drew}, {Barentsen}, \& {Eisl{\"o}ffel}}]{Macfarlane2015}
{Macfarlane}, S.~A., {Toma}, R., {Ramsay}, G., {et~al.} 2015, \mnras, 454, 507

\bibitem[{{McInnes} {et~al.}(2018){McInnes}, {Healy}, \& {Melville}}]{McInnes2018}
{McInnes}, L., {Healy}, J., \& {Melville}, J. 2018, arXiv e-prints, arXiv:1802.03426

\bibitem[{McInnes {et~al.}(2018)McInnes, Healy, Saul, \& Grossberger}]{umap-software2018}
McInnes, L., Healy, J., Saul, N., \& Grossberger, L. 2018, The Journal of Open Source Software, 3, 861

\bibitem[{{Monsalves} {et~al.}(2024){Monsalves}, {Jaque Arancibia}, {Bayo}, {S{\'a}nchez-S{\'a}ez}, {Angeloni}, {Damke}, \& {Segura Van de Perre}}]{Monsalves2024}
{Monsalves}, N., {Jaque Arancibia}, M., {Bayo}, A., {et~al.} 2024, \aap, 691, A106

\bibitem[{{Morales-Rueda} {et~al.}(2006){Morales-Rueda}, {Groot}, {Augusteijn}, {Nelemans}, {Vreeswijk}, \& {van den Besselaar}}]{Morales-Rueda2006}
{Morales-Rueda}, L., {Groot}, P.~J., {Augusteijn}, T., {et~al.} 2006, \mnras, 371, 1681

\bibitem[{{Ostrowski} {et~al.}(2021){Ostrowski}, {Baran}, {Sanjayan}, \& {Sahoo}}]{Ostrowski2021}
{Ostrowski}, J., {Baran}, A.~S., {Sanjayan}, S., \& {Sahoo}, S.~K. 2021, \mnras, 503, 4646

\bibitem[{{Pantoja} {et~al.}(2022){Pantoja}, {Catelan}, {Pichara}, \& {Protopapas}}]{Pantoja2022}
{Pantoja}, R., {Catelan}, M., {Pichara}, K., \& {Protopapas}, P. 2022, \mnras, 517, 3660

\bibitem[{Pedregosa {et~al.}(2011)Pedregosa, Varoquaux, Gramfort, Michel, Thirion, Grisel, Blondel, Prettenhofer, Weiss, Dubourg, Vanderplas, Passos, Cournapeau, Brucher, Perrot, \& Duchesnay}]{scikit-learn}
Pedregosa, F., Varoquaux, G., Gramfort, A., {et~al.} 2011, Journal of Machine Learning Research, 12, 2825

\bibitem[{{Pelisoli} {et~al.}(2020){Pelisoli}, {Vos}, {Geier}, {Schaffenroth}, \& {Baran}}]{Pesoli2020}
{Pelisoli}, I., {Vos}, J., {Geier}, S., {Schaffenroth}, V., \& {Baran}, A.~S. 2020, \aap, 642, A180

\bibitem[{{Pietrukowicz} {et~al.}(2017){Pietrukowicz}, {Dziembowski}, {Latour}, {Angeloni}, {Poleski}, {di Mille}, {Soszy{\'n}ski}, {Udalski}, {Szyma{\'n}ski}, {Wyrzykowski}, {Koz{\l}owski}, {Skowron}, {Skowron}, {Mr{\'o}z}, {Pawlak}, \& {Ulaczyk}}]{Pietrukowicz2017}
{Pietrukowicz}, P., {Dziembowski}, W.~A., {Latour}, M., {et~al.} 2017, Nature Astronomy, 1, 0166

\bibitem[{{Pojmanski}(2002)}]{Pojmanski2002}
{Pojmanski}, G. 2002, \actaa, 52, 397

\bibitem[{{Ranaivomanana} {et~al.}(2023){Ranaivomanana}, {Johnston}, {Groot}, {Aerts}, {Lees}, {IJspeert}, {Bloemen}, {Klein-Wolt}, {Woudt}, {K{\"o}rding}, {Le Poole}, \& {Pieterse}}]{Ranaivomanana2023}
{Ranaivomanana}, P., {Johnston}, C., {Groot}, P.~J., {et~al.} 2023, \aap, 672, A69

\bibitem[{{Reed} {et~al.}(2020){Reed}, {Shoaf}, {N{\'e}meth}, {Vos}, {Uzundag}, {Baran}, {Sahoo}, {Jeffery}, {Telting}, \& {{\O}stensen}}]{Reed2020}
{Reed}, M.~D., {Shoaf}, K.~A., {N{\'e}meth}, P., {et~al.} 2020, \mnras, 493, 5162

\bibitem[{{Richards} {et~al.}(2012){Richards}, {Starr}, {Miller}, {Bloom}, {Butler}, {Brink}, \& {Crellin-Quick}}]{Richards2012}
{Richards}, J.~W., {Starr}, D.~L., {Miller}, A.~A., {et~al.} 2012, \apjs, 203, 32

\bibitem[{{Ricker} {et~al.}(2015){Ricker}, {Winn}, {Vanderspek}, {Latham}, {Bakos}, {Bean}, {Berta-Thompson}, {Brown}, {Buchhave}, {Butler}, {Butler}, {Chaplin}, {Charbonneau}, {Christensen-Dalsgaard}, {Clampin}, {Deming}, {Doty}, {De Lee}, {Dressing}, {Dunham}, {Endl}, {Fressin}, {Ge}, {Henning}, {Holman}, {Howard}, {Ida}, {Jenkins}, {Jernigan}, {Johnson}, {Kaltenegger}, {Kawai}, {Kjeldsen}, {Laughlin}, {Levine}, {Lin}, {Lissauer}, {MacQueen}, {Marcy}, {McCullough}, {Morton}, {Narita}, {Paegert}, {Palle}, {Pepe}, {Pepper}, {Quirrenbach}, {Rinehart}, {Sasselov}, {Sato}, {Seager}, {Sozzetti}, {Stassun}, {Sullivan}, {Szentgyorgyi}, {Torres}, {Udry}, \& {Villasenor}}]{Ricker2015}
{Ricker}, G.~R., {Winn}, J.~N., {Vanderspek}, R., {et~al.} 2015, Journal of Astronomical Telescopes, Instruments, and Systems, 1, 014003

\bibitem[{{Ritter} \& {Kolb}(2003)}]{Ritter2003}
{Ritter}, H. \& {Kolb}, U. 2003, \aap, 404, 301

\bibitem[{Rousseeuw(1987)}]{Rousseeuw1987}
Rousseeuw, P.~J. 1987, Journal of Computational and Applied Mathematics, 20, 53

\bibitem[{{Saffer} {et~al.}(1994){Saffer}, {Bergeron}, {Koester}, \& {Liebert}}]{Saffer1994}
{Saffer}, R.~A., {Bergeron}, P., {Koester}, D., \& {Liebert}, J. 1994, \apj, 432, 351

\bibitem[{{Saha} \& {Vivas}(2017)}]{Saha2017}
{Saha}, A. \& {Vivas}, A.~K. 2017, \aj, 154, 231

\bibitem[{{Sahoo} {et~al.}(2020){Sahoo}, {Baran}, {Heber}, {Ostrowski}, {Sanjayan}, {Silvotti}, {Irrgang}, {Uzundag}, {Reed}, {Shoaf}, {Raddi}, {Vuckovic}, {Ghasemi}, {Zong}, \& {Bell}}]{Sahoo2020}
{Sahoo}, S.~K., {Baran}, A.~S., {Heber}, U., {et~al.} 2020, \mnras, 495, 2844

\bibitem[{{Scargle}(1982)}]{Scargle1982}
{Scargle}, J.~D. 1982, \apj, 263, 835

\bibitem[{{Schaffenroth} {et~al.}(2019){Schaffenroth}, {Barlow}, {Geier}, {Vu{\v{c}}kovi{\'c}}, {Kilkenny}, {Wolz}, {Kupfer}, {Heber}, {Drechsel}, {Kimeswenger}, {Marsh}, {Wolf}, {Pelisoli}, {Freudenthal}, {Dreizler}, {Kreuzer}, \& {Ziegerer}}]{Schaffenroth2019}
{Schaffenroth}, V., {Barlow}, B.~N., {Geier}, S., {et~al.} 2019, \aap, 630, A80

\bibitem[{{Schaffenroth} {et~al.}(2023){Schaffenroth}, {Barlow}, {Pelisoli}, {Geier}, \& {Kupfer}}]{Schaffenroth2023}
{Schaffenroth}, V., {Barlow}, B.~N., {Pelisoli}, I., {Geier}, S., \& {Kupfer}, T. 2023, \aap, 673, A90

\bibitem[{{Schaffenroth} {et~al.}(2022){Schaffenroth}, {Pelisoli}, {Barlow}, {Geier}, \& {Kupfer}}]{Schaffenroth2022}
{Schaffenroth}, V., {Pelisoli}, I., {Barlow}, B.~N., {Geier}, S., \& {Kupfer}, T. 2022, \aap, 666, A182

\bibitem[{Schwarzenberg-Czerny(1996)}]{Schwarzenberg-Czerny1996}
Schwarzenberg-Czerny, A. 1996, The Astrophysical Journal, 460, L107

\bibitem[{{Silvotti} {et~al.}(2022){Silvotti}, {N{\'e}meth}, {Telting}, {Baran}, {{\O}stensen}, {Ostrowski}, {Sahoo}, \& {Prins}}]{Silvotti2022}
{Silvotti}, R., {N{\'e}meth}, P., {Telting}, J.~H., {et~al.} 2022, \mnras, 511, 2201

\bibitem[{{Steeghs} {et~al.}(2022){Steeghs}, {Galloway}, {Ackley}, {Dyer}, {Lyman}, {Ulaczyk}, {Cutter}, {Mong}, {Dhillon}, {O'Brien}, {Ramsay}, {Poshyachinda}, {Kotak}, {Nuttall}, {Pall{\'e}}, {Breton}, {Pollacco}, {Thrane}, {Aukkaravittayapun}, {Awiphan}, {Burhanudin}, {Chote}, {Chrimes}, {Daw}, {Duffy}, {Eyles-Ferris}, {Gompertz}, {Heikkil{\"a}}, {Irawati}, {Kennedy}, {Killestein}, {Kuncarayakti}, {Levan}, {Littlefair}, {Makrygianni}, {Marsh}, {Mata-Sanchez}, {Mattila}, {Maund}, {McCormac}, {Mkrtichian}, {Mullaney}, {Noysena}, {Patel}, {Rol}, {Sawangwit}, {Stanway}, {Starling}, {Str{\o}m}, {Tooke}, {West}, {White}, \& {Wiersema}}]{Steeghs2022}
{Steeghs}, D., {Galloway}, D.~K., {Ackley}, K., {et~al.} 2022, \mnras, 511, 2405

\bibitem[{{Uzundag} {et~al.}(2021){Uzundag}, {C{\'o}rsico}, {Kepler}, {Althaus}, {Werner}, {Reindl}, {Bell}, {Higgins}, {da Rosa}, {Vu{\v{c}}kovi{\'c}}, \& {Istrate}}]{Uzundag2021}
{Uzundag}, M., {C{\'o}rsico}, A.~H., {Kepler}, S.~O., {et~al.} 2021, \aap, 655, A27

\bibitem[{{Uzundag} {et~al.}(2024){Uzundag}, {Krzesinski}, {Pelisoli}, {N{\'e}meth}, {Silvotti}, {Vu{\v{c}}kovi{\'c}}, {Dawson}, \& {Geier}}]{Uzundag2024}
{Uzundag}, M., {Krzesinski}, J., {Pelisoli}, I., {et~al.} 2024, \aap, 684, A118

\bibitem[{{Uzundag} {et~al.}(2023){Uzundag}, {Silvotti}, {Baran}, {Vu{\v{c}}kovi{\'c}}, {N{\'e}meth}, {Sahoo}, \& {Reed}}]{Uzundag2023}
{Uzundag}, M., {Silvotti}, R., {Baran}, A.~S., {et~al.} 2023, Bulletin de la Societe Royale des Sciences de Liege, 92, 11294

\bibitem[{van~der Maaten \& Hinton(2008)}]{Vandermaaten2008}
van~der Maaten, L. \& Hinton, G. 2008, Journal of Machine Learning Research, 9, 2579

\bibitem[{{Van Grootel} {et~al.}(2010){Van Grootel}, {Charpinet}, {Fontaine}, {Green}, \& {Brassard}}]{VanGrootel2010}
{Van Grootel}, V., {Charpinet}, S., {Fontaine}, G., {Green}, E.~M., \& {Brassard}, P. 2010, \aap, 524, A63

\bibitem[{{VanderPlas}(2018)}]{Vanderplas2018}
{VanderPlas}, J.~T. 2018, \apjs, 236, 16

\bibitem[{{Vos} {et~al.}(2020){Vos}, {Bobrick}, \& {Vu{\v{c}}kovi{\'c}}}]{Vos2020}
{Vos}, J., {Bobrick}, A., \& {Vu{\v{c}}kovi{\'c}}, M. 2020, \aap, 641, A163

\bibitem[{{Vos} {et~al.}(2019){Vos}, {Vu{\v{c}}kovi{\'c}}, {Chen}, {Han}, {Boudreaux}, {Barlow}, {{\O}stensen}, \& {N{\'e}meth}}]{vos2019}
{Vos}, J., {Vu{\v{c}}kovi{\'c}}, M., {Chen}, X., {et~al.} 2019, \mnras, 482, 4592

\bibitem[{{Webbink}(1984)}]{Webbink1984}
{Webbink}, R.~F. 1984, \apj, 277, 355

\end{thebibliography}
\begin{appendix}
\onecolumn
\section{Additional material}\label{appendix:A}
\renewcommand{\arraystretch}{0.9}
\setlength{\tabcolsep}{9pt}
\small
\begin{longtable}{clc}
\caption{ All 84 features used in the feature selection.
}\label{tab:all_features} \\ \hline \multicolumn{1}{l}{No.}  & \multicolumn{1}{l}{Feature}&\multicolumn{1}{c}{Description}\\ \hline 
\endfirsthead

\multicolumn{3}{c}%
{\tablename\ \thetable{}. Continued.} \\
\hline \multicolumn{1}{l}{No.}  & \multicolumn{1}{l}\textbf{Feature}&\multicolumn{1}{c}\textbf{Description}\\ \hline 
\endhead

\hline \multicolumn{3}{r}{{Continued on next page}} \\ \hline
\endfoot

\hline
\endlastfoot

\hline \hline

&&Selected features for the clustering analysis\\ \hline 
1       &       log\_sigvar*    &       Significance of variability in the G band in a log scale\\                         
2       &       frac\_period*   &       Period over the standard deviation (std) of the three band {\it Gaia} lightcurve periods\\                         
3       &       std*    &       Standard deviation of the G, BP, and RP periods\\                               
4       &       fapG*   &       False alarm probability of the Lomb-Scargle dominant frequency peak (G band)\\                              
5       &       fapRP*  &       False alarm probability of the Lomb-Scargle dominant frequency peak (RP band)\\                             
6       &       fapBP*  &       False alarm probability of the Lomb-Scargle dominant frequency peak (BP band) \\                            
7       &       Period\_G*      &       Derived period from the G-band lightcurve\\                             
8       &       Period\_BP*     &       Derived period in the BP-band lightcurve\\                              
9       &       period\_RP*     &       Derived period in the RP-band lightcurve\\                              
10      &       amp\_G* &       Amplitude of variability in the G band (mag.)\\                         
11      &       amp\_BP*        &       Amplitude of variability in the BP band (mag.)\\                           
12      &       kurtosisG*      &       G-band kurtosis of the periodogram\\                            
13      &       p99*    &       99th percentile of all periodogram peaks based on the G-band lightcurves\\                               
14      &       p95\_100*       &       95th percentile of the first 100 frequency peaks based on the G-band lightcurves\\                               
15      &       n05*    &       Number of peaks above 0.5 of the normalised $\Psi-$ periodogram based on the G band\\                               
16      &       psi\_sigvar*    &       G-band median absolute deviation of the periodogram\\    
17      &       bp\_rp $\nmid$  &       BP$-$RP colour\\                                
18      &       range\_mag\_g\_fov      &       The range of the G-band time series\\                                
19      &       abbe\_mag\_g\_fov       &       The Abbe value of the G-band time series \\                          
20      &       iqr\_mag\_g\_fov        &       The Interquartile Range (IQR) of the G-band time series\\                             
21      &       mad\_mag\_g\_fov        &       The Median Absolute Deviation (MAD) of the G-band time series\\                               
22      &       stetson\_mag\_g\_fov    &       The single-band Stetson variability index \\                                
23      &       abbe\_mag\_bp   &       The Abbe value of the BP-band time series\\                                
24      &       abbe\_mag\_rp   &       The Abbe value of the RP-band time series\\                                
25      &       outlier\_median\_g\_fov &       Greatest absolute deviation from the G median normalised by the error\\                             
26      &       skewness\_mag\_bp       &       The standardised unbiased unweighted skewness of the BP-band time series\\                                
27      &       std\_dev\_over\_rms\_err\_mag\_g\_fov   &       S/N ratio G estimate\\    \hline \hline
&&Excluded features in the feature selection processes\\
\hline 
28      &       G\_abs* &       Gaia G absolute magnitude\\                             
29      &       N\_G*   &       Number of observations in the G band.\\                         
30      &       N\_BP*  &       Number of observations in the BP band\\                         
31      &       N\_RP*  &       Number of observations in the RP band.\\                                
32      &       amp\_RP*        &       Amplitude of variability in the RP band (mag.)\\                           
33      &       p90\_100*       &       90th percentile of the first 100 frequency peaks\\                               
34      &       p99\_100*       &       99th percentile of the first 100 frequency peaks\\                               
35      &       rmse\_over\_ptp\_amp*   &       Root mean square error (RMSE) of the Lomb-Scargle model \\                            
        &               &       fit over the peak-to-peak G amplitude           \\              
36      &       parallax $\nmid$        &       Gaia parallax\\                         
37      &       parallax\_error $\nmid$ &       Gaia parallax error\\                           
38      &       phot\_g\_mean\_mag $\nmid$      &       G-band mean magnitude\\                         
39      &       phot\_g\_n\_obs $\nmid$ &       Number of observation contributing to G photometry\\                               
40      &       RUWE $\nmid$    &       Renormalised unit weight error\\                                
41      &       num\_selected\_g\_fov   &       Total number of G FOV transits selected for variability analysis\\                             
42      &       mean\_obs\_time\_g\_fov &       Mean observation time for G observations\\                                
43      &       time\_duration\_g\_fov  &       Time duration of the G time series\\                                
44      &       min\_mag\_g\_fov        &       The minimum value of the G-band time series\\                            
45      &       max\_mag\_g\_fov        &       The maximum value of the G-band time series\\                            
46      &       mean\_mag\_g\_fov       &       The mean of the G-band time series\\                                
47      &       median\_mag\_g\_fov     &       The median of the G-band time series\\                           
48      &       trimmed\_range\_mag\_g\_fov     &       Trimmed difference between the highest and lowest G-band time series\\                             
49      &       std\_dev\_mag\_g\_fov   &       Square root of the unweighted G magnitude variance\\                          
50      &       skewness\_mag\_g\_fov   &       The standardised unbiased unweighted skewness of the G-band time series\\                         
51      &       kurtosis\_mag\_g\_fov   &       The standardised unbiased unweighted kurtosis of the G-band time series \\                                
52      &       num\_selected\_bp       &       Total number of BP observations selected for variability analysis\\                             
53      &       mean\_obs\_time\_bp     &       Mean observation time for BP observations\\                               
54      &       time\_duration\_bp      &       Time duration of the BP time series\\                                
55      &       min\_mag\_bp    &       The minimum value of the BP-band time series \\                          
56      &       max\_mag\_bp    &       The maximum value of the BP-band time series\\                           
57      &       mean\_mag\_bp   &       The mean of the BP-band time series \\                              
58      &       median\_mag\_bp &       The median of the BP-band time series\\                         
59      &       range\_mag\_bp  &       The range of the BP-band time series\\                          
60      &       trimmed\_range\_mag\_bp &       Trimmed difference between the highest and lowest BP-band time series\\                            
61      &       std\_dev\_mag\_bp       &       Square root of the unweighted BP magnitude variance\\                         
62      &       kurtosis\_mag\_bp       &       The standardised unbiased unweighted kurtosis of the BP-band time series\\                                
63      &       mad\_mag\_bp    &       The Median Absolute Deviation (MAD) of the BP-band time series\\                            
64      &       iqr\_mag\_bp    &       The Interquartile Range (IQR) of the BP-band time series\\                               
65      &       stetson\_mag\_bp        &       The single-band Stetson variability index\\                         
66      &       std\_dev\_over\_rms\_err\_mag\_bp       &       S/N ratio BP estimate\\                           
67      &       outlier\_median\_bp     &       Greatest absolute deviation from the BP median normalised by the error\\                            
68      &       num\_selected\_rp       &       Total number of RP observations selected for variability analysis\\                             
69      &       mean\_obs\_time\_rp     &       Mean observation time for RP observations\\                               
70      &       time\_duration\_rp      &       Time duration of the RP time series\\                                
71      &       min\_mag\_rp    &       The minimum value of the RP-band time series\\                           
72      &       max\_mag\_rp    &       The maximum value of the RP-band time series\\                           
73      &       mean\_mag\_rp   &       The mean of the RP-band time series\\                           
74      &       median\_mag\_rp &       The median of the RP-band time series\\                         
75      &       range\_mag\_rp  &       The range of the RP-band time series\\                          
76      &       trimmed\_range\_mag\_rp &       Trimmed difference between the highest and lowest RP-band time series\\                            
77      &       std\_dev\_mag\_rp       &       Square root of the unweighted RP magnitude variance\\                         
78      &       skewness\_mag\_rp       &       The standardised unbiased unweighted skewness of the RP-band time series\\                                
79      &       kurtosis\_mag\_rp       &       The standardised unbiased unweighted kurtosis of the RP-band time series\\                                
80      &       mad\_mag\_rp    &       The Median Absolute Deviation (MAD) of the RP-band time series\\                            
81      &       iqr\_mag\_rp    &       The Interquartile Range (IQR) of the RP-band time series \\                              
82      &       stetson\_mag\_rp        &       The single-band Stetson variability index\\                         
83      &       std\_dev\_over\_rms\_err\_mag\_rp       &       S/N ratio RP estimate\\                           
84      &       outlier\_median\_rp     &       Greatest absolute deviation from the RP median normalised by the error\\    
\end{longtable}
\tablefoot{ Features marked with (*) were computed in this work, those with ($\nmid$) are from the Gaia DR3 source database \citep{GaiaCollab2023}, while the rest were obtained from the {\it Gaia} variability summary table \citep{Eyer2023}. A full description of these {\it Gaia} statistics can be found in the {\it Gaia} documentation \href{https://gea.esac.esa.int/archive/documentation/GDR3/Gaia_archive/chap_datamodel/sec_dm_variability_tables/ssec_dm_vari_summary.html}{here}.}

\begin{table*}[!h]
    \centering
    \caption{Feature ranking based on the manual labelling.} \citep{Eyer2023}\label{tab:feature_ranking_manual_lab} 
\small
\begin{tabular}{clc}
\toprule
ID & Feature & Description
\\   \midrule

1 & p95\_100* & 95th percentile of the first 100 frequency peaks based on the G-band lightcurves\\
2 & n05* & Number of peaks above 0.5 of the normalised $\Psi$ periodogram based on the G band\\
3 & p99*& 99th percentile of all periodogram peaks based on the G-band lightcurves\\
4 & Period\_G*& Derived period from the G-band lightcurve\\
5 & frac\_period*& Period over the standard deviation (std) of the three band Gaia lightcurve periods\\
6 & fapG*& False alarm probability of the Lomb-Scargle dominant frequency peak (G band)\\
7 & psi\_sigvar*& G-band median absolute deviation of the periodogram  \\
8 & kurtosisG*& G-band kurtosis of the periodogram \\
9 & iqr\_mag\_g\_fov& The Interquartile Range (IQR) of the G-band time series\\
10 & std*& Standard deviation of the G, BP, and RP periods\\
11 & amp\_G*& Amplitude of variability in the G band (mag.)\\
12 & log\_sigvar*& Significance of variability in the G band in a log scale\\
13 & mad\_mag\_g\_fov&The Median Absolute Deviation (MAD) of the G-band time series \\
14 & range\_mag\_g\_fov& The range of the G-band time series\\
15 & abbe\_mag\_bp&The Abbe value of the BP-band time series \\
16 & abbe\_mag\_rp&The Abbe value of the RP-band time series \\
17 & Period\_RP*& Derived period from the RP-band lightcurve\\
18 & fapBP*& False alarm probability of the Lomb-Scargle dominant frequency peak (BP band)\\
19 & abbe\_mag\_g\_fov& The Abbe value of the G-band time series\\
20 & stetson\_mag\_g\_fov& Stetson G FoV variability index\\
21 & Period\_BP*& Derived period from the BP-band lightcurve\\
22 & amp\_BP*& Amplitude of variability in the BP band(mag.)\\
23 & fapRP*& False alarm probability of the Lomb-Scargle dominant frequency peak (RP band)\\
24 & std\_dev\_over\_rms\_err\_mag\_g\_fov&S/N ratio G FoV estimate \\
25 & bp\_rp $\nmid$& BP $-$ RP colour\\
26 & outlier\_median\_g\_fov& The most outlying measurement with respect to the median \\
27 & skewness\_mag\_bp& The standardised unbiased unweighted skewness of the BP-band time series\\
\bottomrule 
\end{tabular}
\tablefoot{
Features marked with (*) were computed in this work,  those with ($\nmid$) are from the {\it Gaia} DR3 source database \citep{GaiaCollab2023}, while the rest were obtained from the {\it Gaia} variability summary table.}
\end{table*}
\begin{table*}[h!]
    \centering
    \caption{Feature ranking based on the three cluster labels.}\label{tab:feature_ranking_clusters} 
\small
\begin{tabular}{clc}
\toprule
ID & Feature & Description
\\   \midrule
1 & amp\_G*& Amplitude of variability in the G band (mag.)\\
2 & range\_mag\_g\_fov& The range of the G-band time series\\
3 & iqr\_mag\_g\_fov& The Interquartile Range (IQR) of the G-band time series\\
4 & log\_sigvar*& Significance of variability in the G band in a log scale \\
5 & stetson\_mag\_g\_fov& Stetson G FoV variability index\\
6 & n05*& Number of peaks above 0.5 of the normalised $\Psi$ periodogram based on the G band\\
7 & std\_dev\_over\_rms\_err\_mag\_g\_fov&S/N ratio G FoV estimate \\
8 & p95\_100*& 95th percentile of the first 100 frequency peaks based on the G-band lightcurves\\
9 & mad\_mag\_g\_fov&The Median Absolute Deviation (MAD) of the G-band time series \\
10 & bp\_rp $\nmid$& BP $-$ RP colour\\
11 & outlier\_median\_g\_fov&The most outlying measurement with respect to the median \\
12 & p99*& 99th percentile of all periodogram peaks based on the G-band lightcurves\\
13 & amp\_BP*& Amplitude of variability in the BP band(mag.)\\
14 & Period\_G*& Derived period from the G-band lightcurve\\
15 & abbe\_mag\_g\_fov& The Abbe value of the G-band time series\\
16 & kurtosisG*& G-band kurtosis of the periodogram \\
17 & skewness\_mag\_bp& The standardised unbiased unweighted skewness of the BP-band time series\\
18 & abbe\_mag\_bp&The Abbe value of the BP-band time series \\
19 & psi\_sigvar*& G-band median absolute deviation of the periodogram  \\
20 & abbe\_mag\_rp&The Abbe value of the RP-band time series \\
21 & fapG*& False alarm probability of the Lomb-Scargle dominant frequency peak (G band)\\
22 & Period\_RP*& Derived period from the RP-band lightcurve\\
23 & Period\_BP*& Derived period from the BP-band lightcurve\\
24 & frac\_period*& Period over the standard deviation (std) of the three band Gaia lightcurve periods\\
25 & std*& Standard deviation of the G, BP, and RP periods\\
26 & fapRP*& False alarm probability of the Lomb-Scargle dominant frequency peak (RP band)\\
27 & fapBP*& False alarm probability of the Lomb-Scargle dominant frequency peak (BP band)\\
\bottomrule \\
\end{tabular}
\tablefoot{Features marked with (*) were computed in this work,  those with ($\nmid$) are from the {\it Gaia} DR3 source database \citep{GaiaCollab2023}, while the rest were obtained from the {\it Gaia} variability summary table \citep{Eyer2023}.}
\end{table*}

\begin{figure*}
  \centering
\begin{subfigure}[b]{0.3\textwidth}
        \centering
\includegraphics[width=\textwidth]{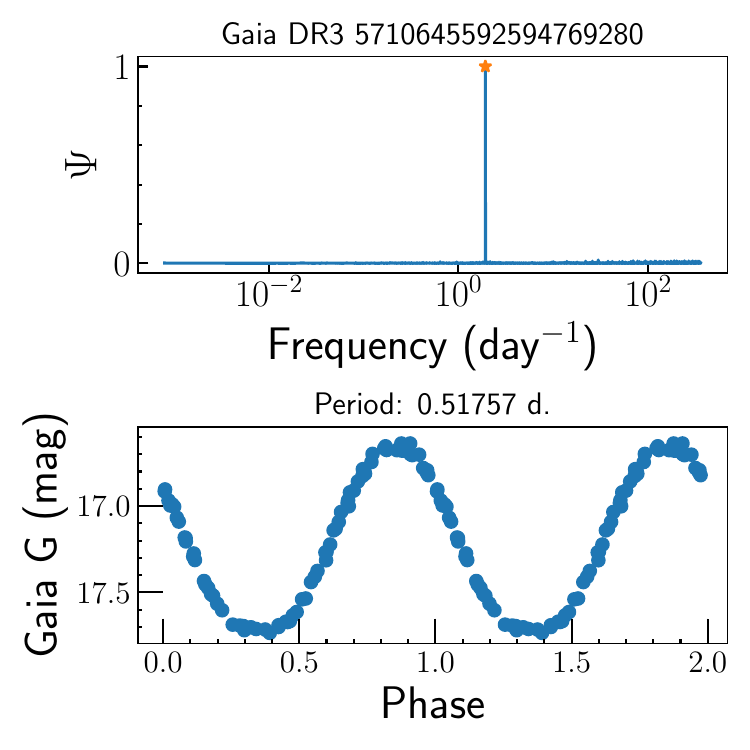}
        \caption{}
         
    \end{subfigure}
    \hfill
    \begin{subfigure}[b]{0.3\textwidth}
        \centering
\includegraphics[width=\textwidth]{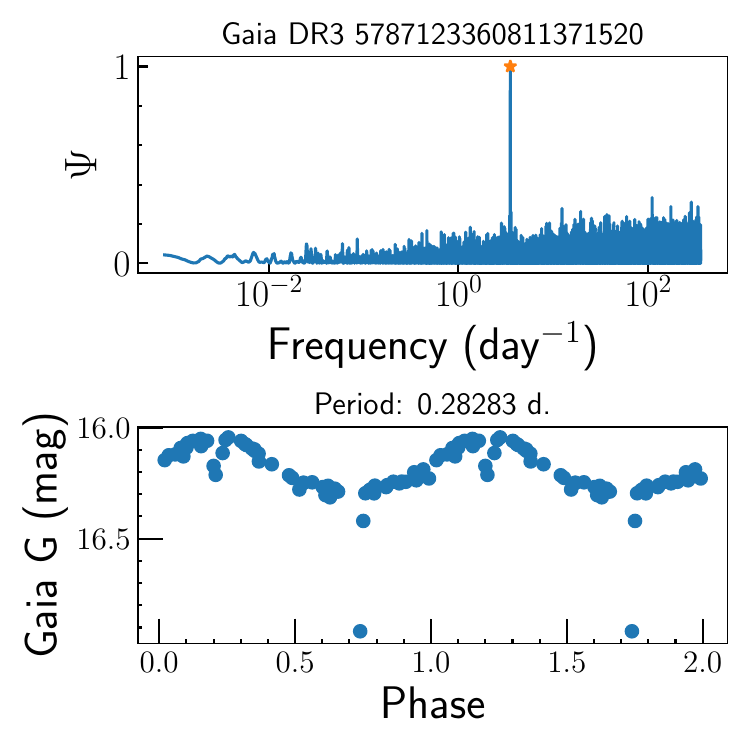}
        \caption{}
          
\end{subfigure}
\hfill
    \begin{subfigure}[b]{0.3\textwidth}
        \centering
\includegraphics[width=\textwidth]{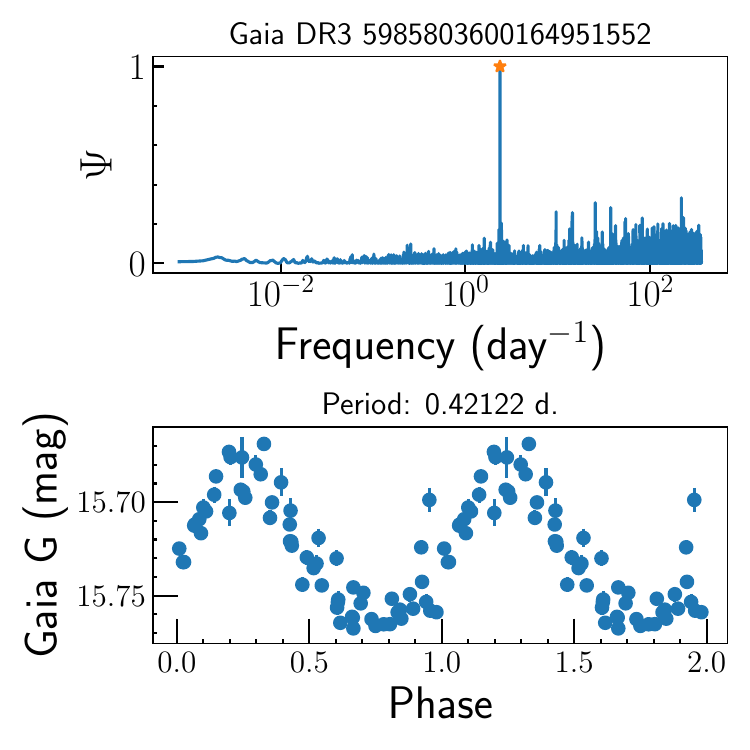}
        \caption{}
          
\end{subfigure}
\caption*{Three examples of objects (a, b, and c) in cluster 0.}

\begin{subfigure}[b]{0.3\textwidth}
        \centering
\includegraphics[width=\textwidth]{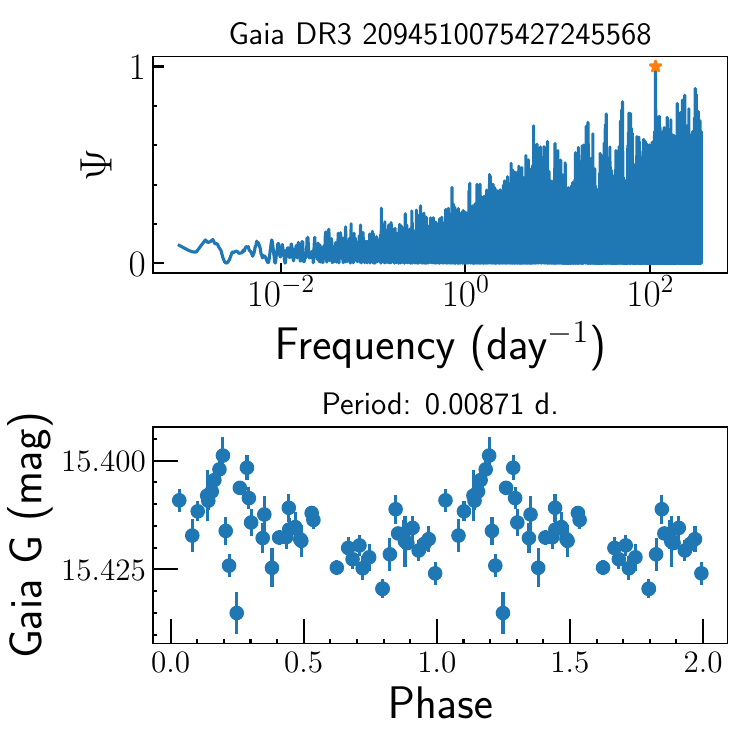}
        \caption{}
         
    \end{subfigure}
    \hfill
    \begin{subfigure}[b]{0.3\textwidth}
        \centering
\includegraphics[width=\textwidth]{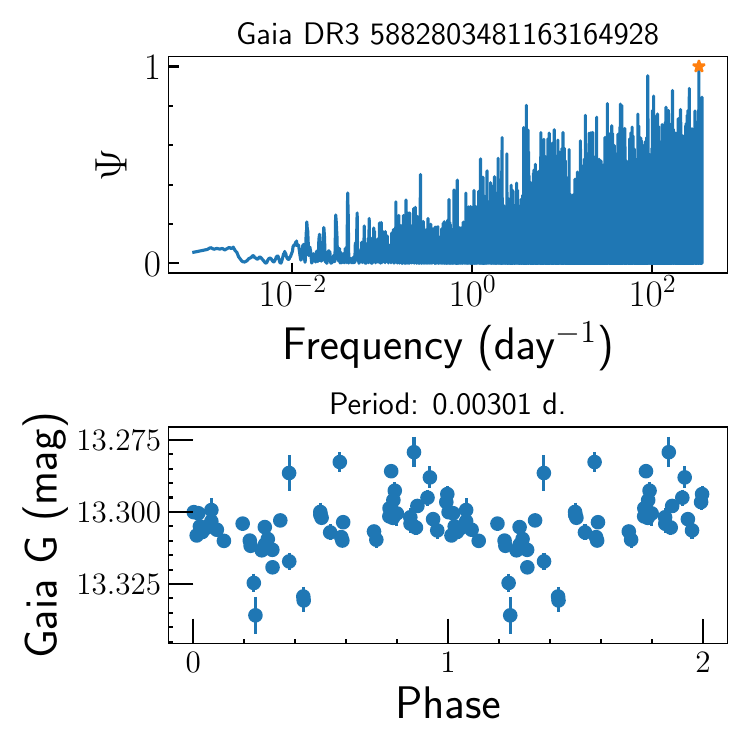}
        \caption{}
          
\end{subfigure}
\hfill
    \begin{subfigure}[b]{0.3\textwidth}
        \centering
\includegraphics[width=\textwidth]{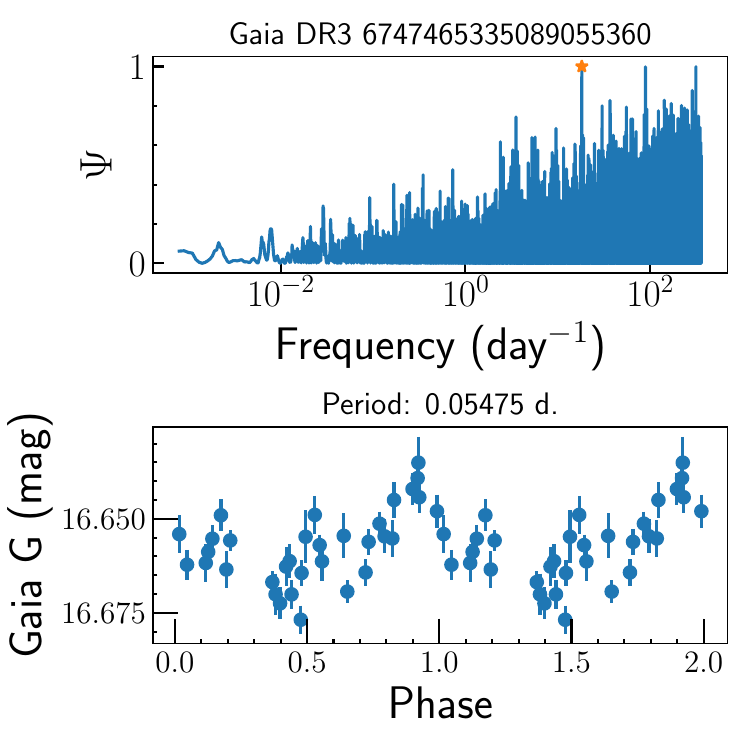}
        \caption{}
          
\end{subfigure}
\caption*{Three examples of objects (d, e, and f) in cluster 1.}

\begin{subfigure}[b]{0.3\textwidth}
        \centering
\includegraphics[width=\textwidth]{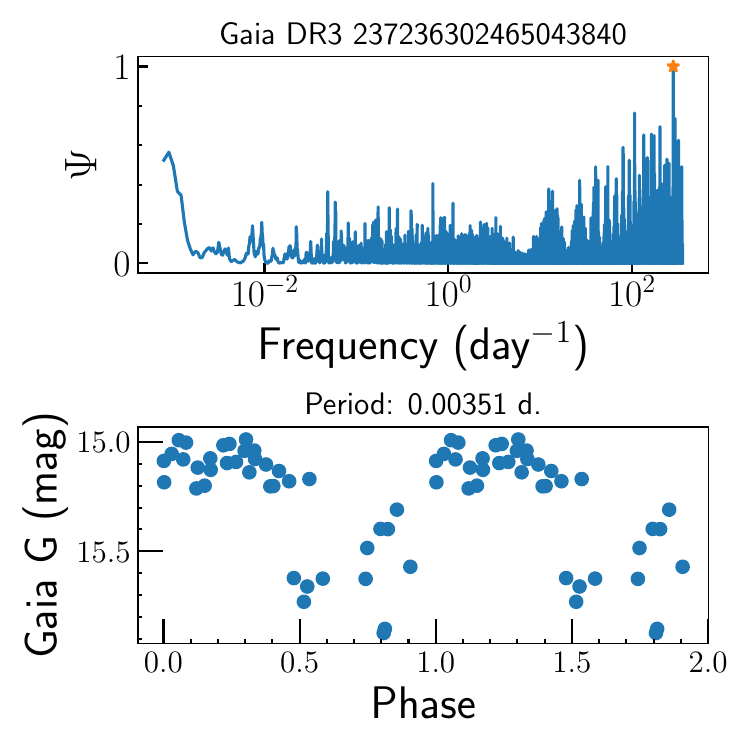}
        \caption{}
         
    \end{subfigure}
    \hfill
    \begin{subfigure}[b]{0.3\textwidth}
        \centering
\includegraphics[width=\textwidth]{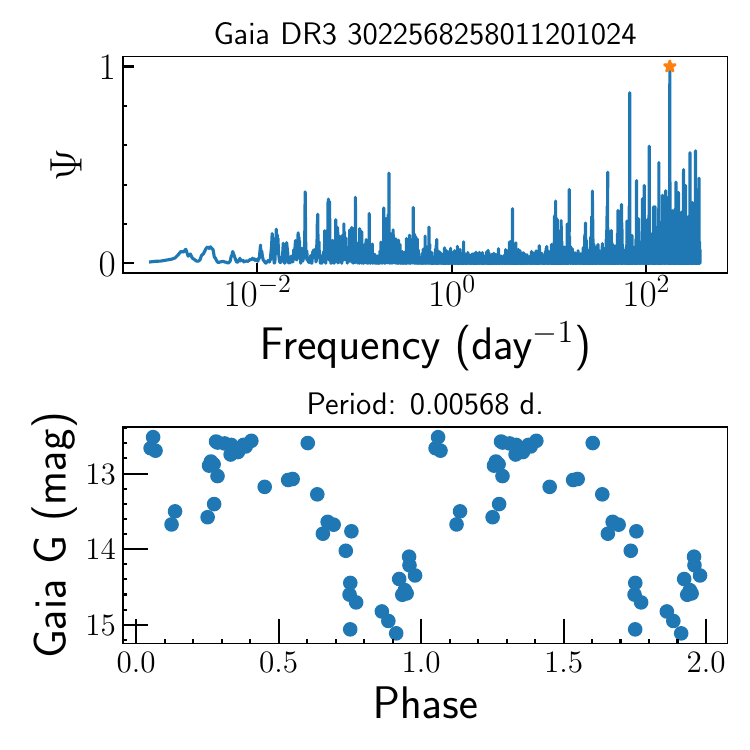}
        \caption{}
          
\end{subfigure}
\hfill
    \begin{subfigure}[b]{0.3\textwidth}
        \centering
\includegraphics[width=\textwidth]{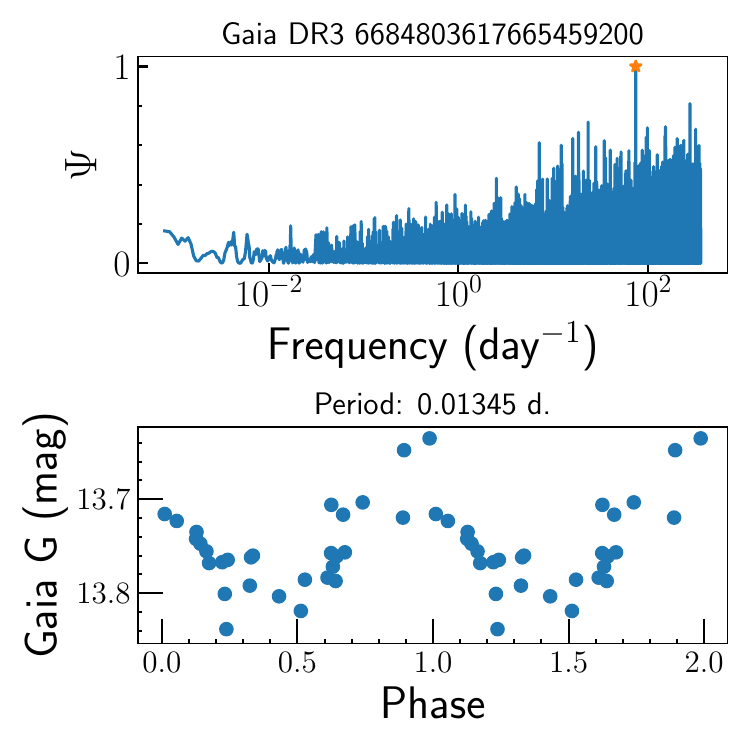}
        \caption{}
          
\end{subfigure}
\caption*{Three examples of objects (g, h, and i) in cluster 2.}

\caption{Examples of periodograms and phase-folded light curves for each cluster. The top, middle, and bottom rows correspond to cluster 0, cluster 1, and cluster 2, respectively.}\label{fig:cluster_lc_examples}

\end{figure*}

\begin{figure*}
  \centering
\includegraphics[width=0.7\linewidth]{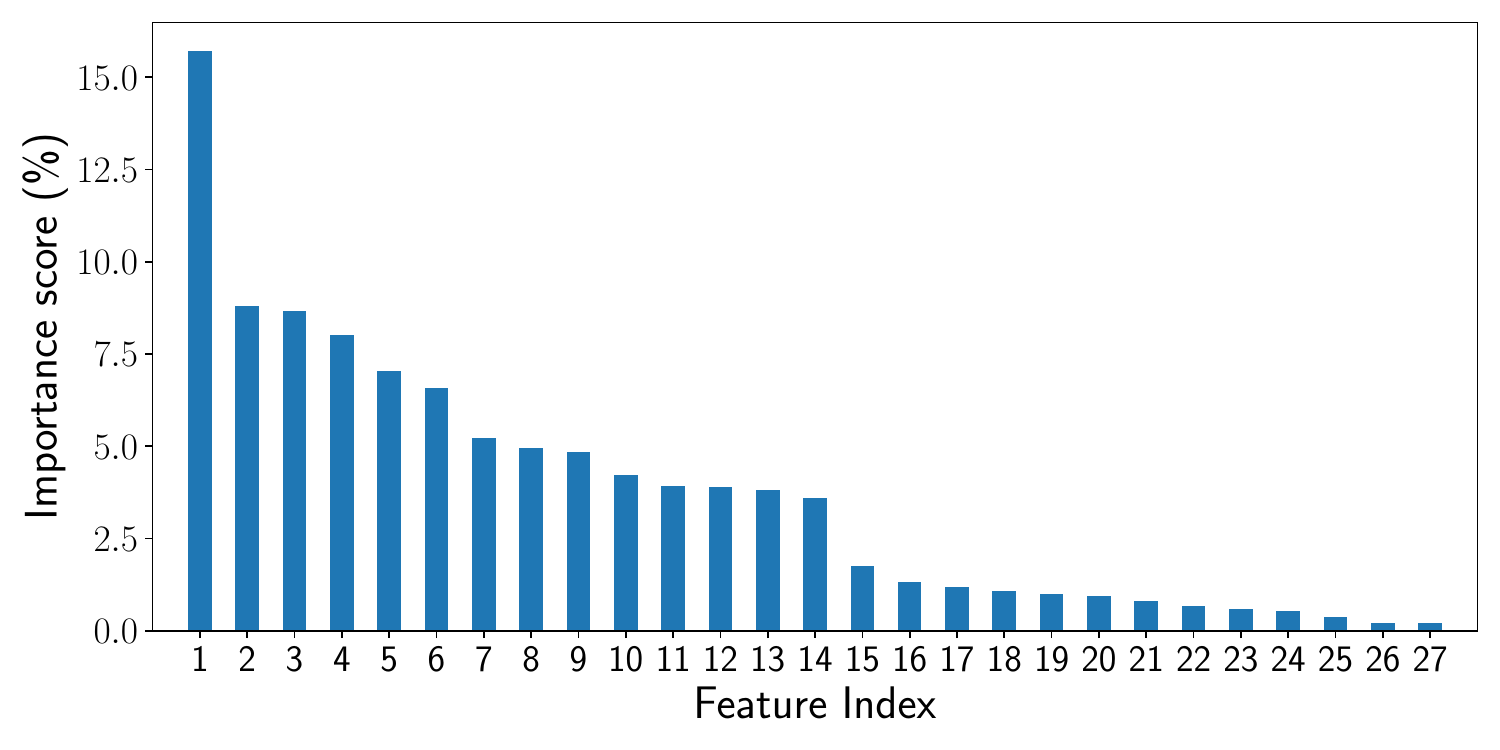}
\caption{Random forest feature importance scores for the 27 features listed in Table \ref{tab:feature_ranking_clusters}. The x-axis corresponds to the Feature ID in the table.}
\label{fig:features_importance_score}
\end{figure*}

\begin{figure*}
  \centering
\includegraphics[width=0.5\linewidth]{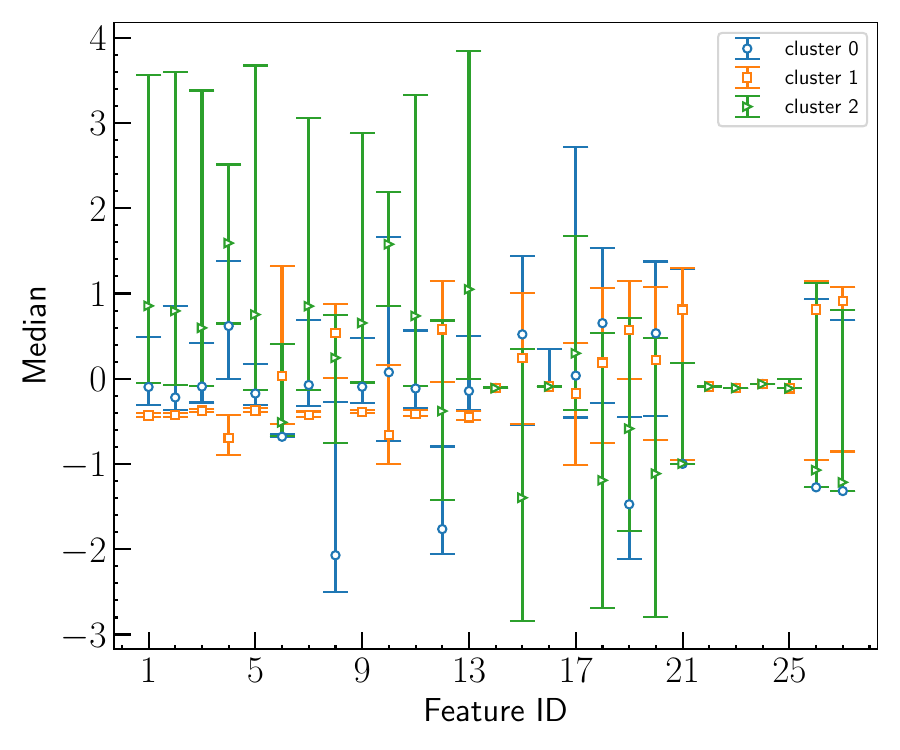}
\caption{Distribution of the feature medians for each cluster. The x-axis corresponds to the feature ID listed in Table \ref{tab:feature_ranking_clusters}; The y-axis represents the median (open marker), the 10th percentile (lower cap), and the 90th percentile (upper cap)  of each feature after a z-score normalisation.}
\label{fig:features_median_iqr}
\end{figure*}

\end{appendix}
\end{document}